\documentclass[%
 reprint,
 amsmath,amssymb,
 aps,
]{revtex4-2}
\usepackage{xcolor}
\usepackage{graphicx}
\usepackage{dcolumn}
\usepackage{bm}

\newcommand{\atan}{\tan^{-1}}
\begin{document}

\preprint{APS/123-QED}


\title{Electronic and optical conductivity of kekul\'e-patterned graphene:\\ Intravalley and intervalley transport}

\author{Sa\'ul A. Herrera}

\author{Gerardo G. Naumis}%
 \email{naumis@fisica.unam.mx}
\affiliation{%
Depto. de Sistemas Complejos, Instituto de F\'isica, \\ Universidad Nacional Aut\'onoma de M\'exico (UNAM)\\
Apdo. Postal 20-364, 01000, CDMX, M\'exico.
}%

\date{\today}

\begin{abstract}
A Kubo formalism is used to calculate the electronic and optical conductivity of a graphene superlattice with Y-shaped kekul\'e bond texture, similar to that visualized in recent experiments. We show that new conduction channels between the valleys in graphene are opened by the kekul\'e distortion. This intervalley contribution to the electronic transport is not present in pristine graphene and here appears due to the folding of the Dirac cones $K$, $K'$ on top of each other. The contribution of intervalley transport to the conductivity of graphene, as well as the modification of the intravalley transport, are analyzed in detail for different frequency, temperature, chemical potential and scattering rate limits. We obtain analytical expressions for the conductivity that reproduce previous expressions used to fit experimental measurements in graphene and compare with direct numerical evaluations of the Kubo formula, finding great agreement. The optical absorption arising from intervalley transitions displays a maximum at a special frequency that can be tuned by doping. Our results show how the single parameter describing the valley coupling in this system could be obtained by measuring graphene's   optical absorbance in the region where interband transitions are blocked by the impurities. Finally, we use Fermi's golden rule to independently verify some of the previous results.
\begin{description}
\item[Key words]
Kekul\'e, graphene, Kubo conductivity
\end{description}
\end{abstract}

\maketitle


\section{\label{sec:level1}Introduction}

Recently, a kekul\'e bond order was found in experimental studies of graphene sheets grown epitaxially on copper substrates \cite{Gutierrez2016}. Later on, a valley-momentum locking was predicted for this system, leading to the emergence of two species of massless Dirac fermions and the removal of degeneracies in the field-dependent Landau levels \cite{Gamayun}. 
Afterwards, it was proved that such system provides a potential research platform for strain-controlled valley-tronics \cite{Elias_2019}. For kekul\'e bond order graphene bilayers,  multiflavor Dirac fermions were predicted to exist \cite{Tijerina_2019}. Also, kekul\'e distortion is one of the suggested mechanisms behind the superconducting behavior in magic angle twisted graphene over graphene \cite{Bitan_2010,Hoi_2019}. Its experimental realization is also  investigated in other kinds of non-atomic systems, as for example in acoustical lattices, where it is possible to produce topological Majorana modes \cite{Penglin_2019}. 
Needless to say, ``artificial" kekul\'e ordering can be produced in polaronic \cite{Cerda2013}, photonic \cite{Photonic_2019} and atomic systems \cite{Atomic_2019}.

It is important to remark that kekul\'e bond order is among one of the most interesting phases resulting from strain in the system \cite{NaumisReview},  showing great promise for applications in the next generation of nanoelectronics \cite{Castro1,NaumisTerrones2009,Sarma}. In general, strain leads to tunable topological quantum phase transitions \cite{Yan2015b,Roman2015} and new interesting topological phase diagrams produced by time-dependent strain fields \cite{Taboada2017,RomanJCP2017}. In fact, one of the first models in which topological phases were observed is the Su-Schrieffer-Heeger model, originally developed to study bond-ordering in polyethylene \cite{Fulde}. Recently, time-dependent bond ordering has been found to produce interesting topological effects  in the   Su-Schrieffer-Heeger model \cite{TimeSSHModel}.

\begin{figure}[b]
\includegraphics[width=0.25\textwidth]{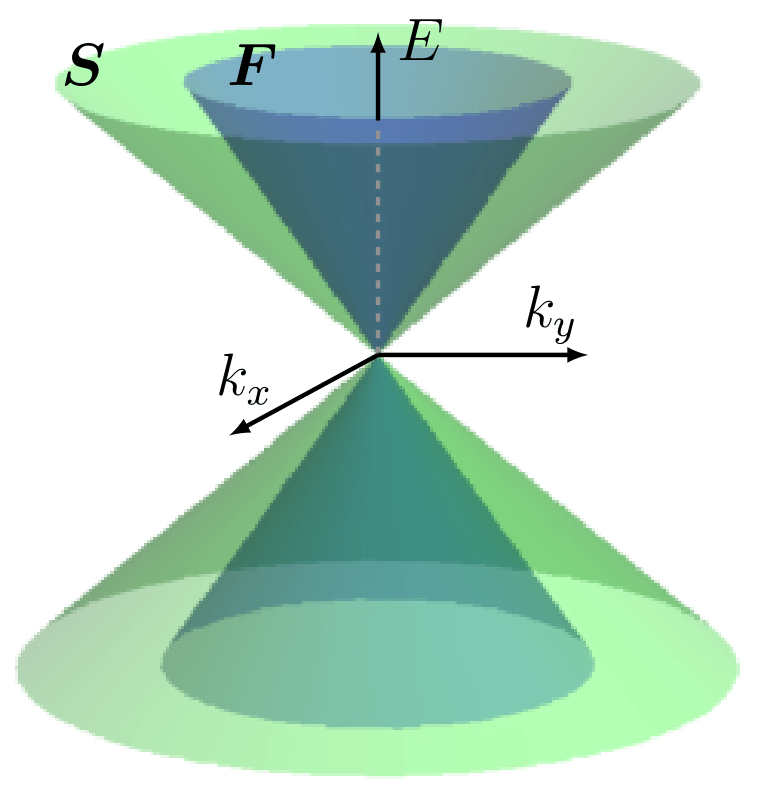}
\caption{\label{Fig:DispersionCone}Energy dispersion of low-energy electronic excitations around the Dirac point in Kek-Y distorted graphene. The letter $S$ labels the slow cone. It has a slope given by the velocity $v_0(1-\Delta_0)$. The letter $F$ labels the fast cone, associated with a velocity $v_0(1+\Delta_0)$}.
\end{figure}

Here we study the consequences of the valley-momentum locking on the electronic transport properties of kekul\'e-distorted graphene using the low-energy Hamiltonian obtained by Gamayun et al. \cite{Gamayun}. We concentrate on the Kek-Y phase, in which the Dirac cones $K$, $K'$ fold on top of each other leading to two species Dirac massless fermions with different velocities (see Fig. \ref{Fig:DispersionCone}).

The corresponding  low-energy  Hamiltonian is given by the following $4\times4$ matrix \cite{Gamayun},

\begin{equation}\label{Eq:HamiltonianMatrix}
    H =  \left( \begin{array}{ccc}
v_0\boldsymbol{p}\cdot\boldsymbol{\sigma} & \tilde{\Delta} Q_\chi\\
\tilde{\Delta}^\ast Q_\chi^\dagger & v_0 \boldsymbol{p}\cdot\boldsymbol{\sigma}  \end{array} \right),
\end{equation}
where $\tilde{\Delta}$ is a coupling amplitude introduced by the bond-density wave that describes the kekul\'e textures  and $\boldsymbol{\sigma}=(\sigma_x,\sigma_y)$ is the set of Pauli matrices. The Kek-Y texture can be described by $Q_\chi=v_0(\chi p_x -i p_y)\sigma_0$ and 
$|\chi|=1$, where $\sigma_0$ is the identity. For simplicity we consider a real $\tilde{\Delta}=\Delta_0$ and $\chi=1$, as  a complex $\tilde{\Delta}$ and $\chi=-1$ are equivalent upon an unitary transformation \cite{Gamayun}. Furthermore, in what follows we will write all the energies in units of $\hbar$ and define the scaled momentum $\boldsymbol{\xi}=v_0 \boldsymbol{k}$. These considerations lead to the Hamiltonian,

\begin{equation}\label{Eq:HamiltonianMatrix2}
    H =  \left( \begin{array}{ccc}
\boldsymbol{\xi}\cdot\boldsymbol{\sigma} & \Delta_0 (\xi_x-i \xi_y)\sigma_0 \\
\Delta_0 (\xi_x+i \xi_y)\sigma_0 &  \boldsymbol{\xi}\cdot\boldsymbol{\sigma}  \end{array} \right),
\end{equation}
or
\begin{equation}\label{Eq:Hamiltonian}
    H=(\boldsymbol{\xi}\cdot\boldsymbol{\sigma})\otimes \tau_0 + \Delta_0 \sigma_0 \otimes (\boldsymbol{\xi}\cdot\boldsymbol{\tau}),
\end{equation}

with $\boldsymbol{\tau}=(\tau_x,\tau_y)$ defining a second pair of Pauli matrices and $\tau_0$ the unitary matrix.

As shown in Fig. \ref{Fig:DispersionCone}, the  kekul\'e distortion folds the  $K$ and $K'$ valleys in graphene into the $\Gamma$ point of the superlattice Brillouin zone, resulting in a ``fast" cone with Fermi velocity $v_0(1+\Delta_0)$ and a ``slow" cone with Fermi velocity $v_0(1-\Delta_0)$. We label these cones as $F$ and $S$, respectively.

\begin{figure}[t]
\includegraphics[width=0.35\textwidth]{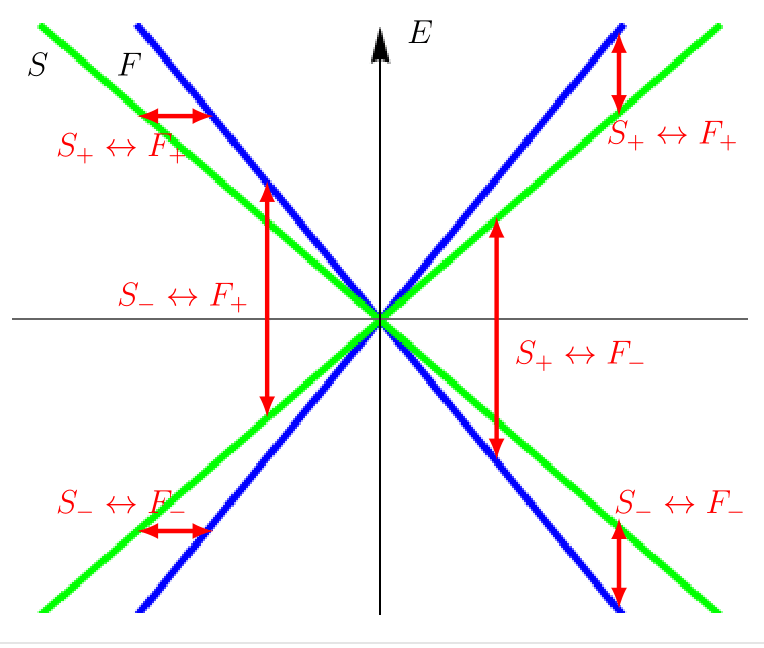}
\caption{\label{Fig:Intervalley_transitions}
Diagram of the different types of intervalley transitions in Kek-Y graphene. Horizontal transitions require scattering aided by disorder while vertical (or optical) transitions are only produced by the field. }
\end{figure}

\begin{figure}[h]
\includegraphics[width=0.35\textwidth]{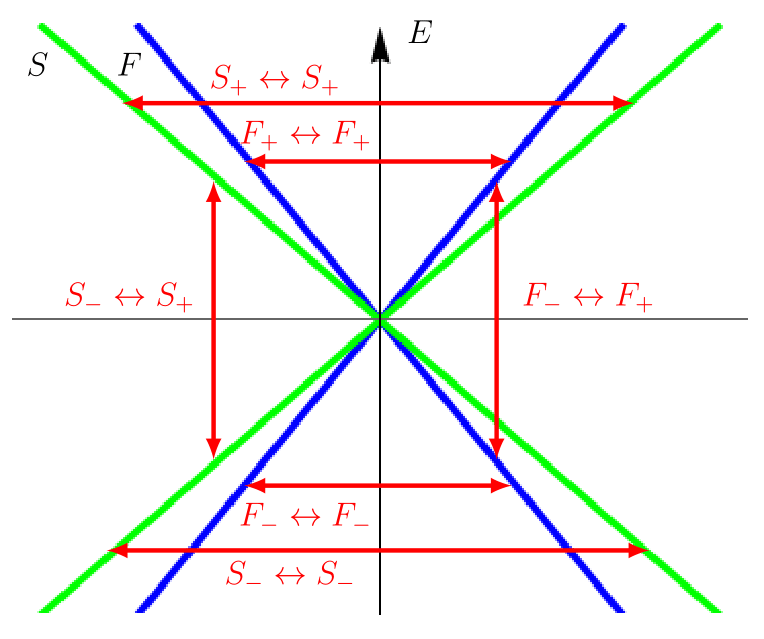}
\caption{\label{Fig:Intravalley_transitions}Diagram of the different types of intravalley transitions in Kek-Y graphene. }
\end{figure}

The aim of this paper is to study the electronic and optical conductivity that results from the Hamiltonian given by Eq. (\ref{Eq:HamiltonianMatrix}). The layout is the following, in sec. \ref{Sec:Conductivity} we compute the  conductivity using a Kubo formalism while in sec. \ref{Sec:Discussion} a discussion is presented concerning the different physical limits of frequency, temperature, scattering rate and chemical potential. A detailed analysis of the different conduction channels is also presented. Finally, the last section is devoted to the conclusions.

\section{Conductivity of Kek-Y distorted graphene}\label{Sec:Conductivity}

To calculate the conductivity resulting from the Hamiltonian given by Eq. (\ref{Eq:HamiltonianMatrix}), we need to find its spectrum and eigenfunctions.
The four eigenvalues of $H$ are given by
\begin{equation}\label{Eq:Dispersion}
    E_{\alpha,\alpha'}=(\alpha + \alpha'\Delta_0)\xi,
\end{equation}
where $\alpha,\alpha'=\pm1$ and $\xi=\sqrt{\xi_x^2 + \xi_y^2}$. The $F$ cone is described by the dispersion $E=\pm (1+\Delta_0)\xi$ and the $S$ cone is described by $E=\pm(1-\Delta_0)\xi$. Here, by ``intervalley transition" we shall mean a transition that implies change in the electron's velocity, from $v_0(1+\Delta_0)$ to $v_0(1-\Delta_0)$, or vice versa. The eigenvectors are,

\begin{equation}
     |\Psi_{\alpha,\alpha'}\rangle= |\Psi_{\alpha}\rangle \otimes |\Psi_{\alpha'}\rangle
\end{equation}
where $|\Psi_{\alpha}\rangle$ are  the eigenvectors for pristine graphene. More explicitly, defining $\theta=\tan^{-1}\xi_y /\xi_x$, the eigenvectors can be written as, 
\begin{equation}
    |\Psi_{\alpha,\alpha'}\rangle=\frac{1}{2}\left( \begin{array}{cccc}
       1,  & \alpha e^{i \theta}, & \alpha' e^{i \theta}, & \alpha\alpha' e^{2 i \theta}
    \end{array}\right)^T.
\end{equation}

We now proceed to find the conductivity. In the Kubo formalism, the real part of the diagonal conductivity is given by \cite{Ziegler,Foa},

\begin{eqnarray}\label{Eq:Sigmadef}
\sigma_{\nu\nu}=\pi \frac{e^2}{\hslash}&&\int \text{Tr}\{[H,r_\nu]\delta(H-\epsilon-\omega)\nonumber\\
&&\times[H,r_\nu]\delta(H-\epsilon)\}\frac{f_\beta(\epsilon+\omega)-f_\beta(\epsilon)}{\omega}d\epsilon
\end{eqnarray}

where $r_{\nu}$ is the position operator in the $\nu=x,y$ direction, $f_\beta(\epsilon)=1/[1+exp(\beta\epsilon)]$ is the Fermi distribution with $\beta=1/k_BT$, $T$ being the temperature and $\delta(x)$ is the Dirac delta function of $x$.

After changing $\epsilon \rightarrow \epsilon - \omega/2$ and defining $\epsilon_\pm=\epsilon\pm\omega/2$ we can write
\begin{equation}\label{Eq:Sigma_2}
    \sigma_{\nu\nu}=-\pi\frac{e^2}{\hbar}\int T(\epsilon)\frac{f_\beta(\epsilon_+)-f_\beta(\epsilon_-)}{\omega}d\epsilon
\end{equation}
where we have expressed the trace in Eq. (\ref{Eq:Sigma_2}) as 
\begin{equation}\label{Eq:Trace_definition}
    T(\epsilon)=\int\sum_{\alpha,\alpha'}\langle\Psi_{\alpha,\alpha'}|\Lambda_+\Lambda_-|\Psi_{\alpha,\alpha'}\rangle\frac{d^2 \xi}{(2\pi)^2},
\end{equation}
obtained by using the Fourier transform of the current operators $-i e [H,r_\nu] \rightarrow e v_0 (\partial H /  \partial \xi_\nu)$ and defining the operators $\Lambda_\pm=(\partial H / \partial \xi_\nu)\delta(H-\epsilon_\pm)$. In what follows we take $\nu=y$ and simply write the diagonal conductivity $\sigma_{yy}$ as $\sigma$, then, from Eq. (\ref{Eq:Hamiltonian}), 
\begin{equation}\label{Eq:Current_op}
    \frac{\partial H}{\partial \xi_y}= \sigma_y \otimes \tau_0 + \Delta_0 \sigma_0 \otimes  \tau_y
\end{equation}

In order to evaluate Eq. (\ref{Eq:Trace_definition}) we rewrite the current operator in the representation of energy eigenstates $|\Psi_{\alpha,\alpha'}\rangle$,
\begin{equation}\label{Eq:Current_op_diag}
  \frac{\partial H}{\partial \xi_y}=\mathcal{U}\sigma_y\mathcal{U^\dagger}\otimes \tau_0 + \Delta_0 \sigma_0 \otimes \mathcal{U}\tau_y  \mathcal{U}^\dagger
\end{equation}
where 
\begin{equation}\label{Eq:U_matrix}
    \mathcal{U}^\dagger=\frac{1}{\sqrt{2}}
    \left( \begin{array}{ccc}
     
    1 & 1 \\
    +e^{i \theta} & -e^{i \theta}
    \end{array} \right)
\end{equation}
so
\begin{equation}\label{Eq:UsigmaUt}
\mathcal{U}\sigma_y\mathcal{U^\dagger}=\mathcal{U}\tau_y\mathcal{U}^\dagger=
\left( \begin{array}{ccc}
    \sin\theta & i \cos \theta  \\
    -i \cos \theta & -\sin \theta
\end{array}\right).
\end{equation}
A scattering rate $\eta$ can be introduced by considering soft Dirac delta functions $\delta_\eta(x)$ in the definitions of the operators $\Lambda_\pm$ given by \cite{Ziegler}
\begin{equation}\label{Eq:Delta_def}
    \delta(x)\approx\delta_\eta(x)=\lim_{\eta\rightarrow 0} \frac{1}{\pi}\frac{\eta}{x^2+\eta^2}
\end{equation}
The parameter $\eta$ can be interpreted as the imaginary part of the self energy introduced by disorder.
Thus, Eq. (\ref{Eq:Trace_definition}) can be written as
\begin{eqnarray}\label{Eq:Trace_expanded}
    T(\epsilon)=\sum_{\alpha,\alpha'}\sum_{\beta,\beta'} \int  && \big\langle\Psi_{\alpha,\alpha'}\big|\frac{\partial H}{\partial \xi_y} \big|\Psi_{\beta,\beta'}\big\rangle   \big\langle\Psi_{\beta,\beta'}\big|\frac{\partial H}{\partial \xi_y}\big|\Psi_{\alpha,\alpha'}\big\rangle \nonumber \\
    && \times \delta_\eta(H-\epsilon_+)\delta_\eta(H-\epsilon_-)\frac{d^2 \xi}{(2\pi)^2}
\end{eqnarray}
with the indexes $\alpha,\alpha',\beta,\beta'=\pm1$. 

The different terms in Eq. (\ref{Eq:Trace_expanded}) are related to the  different types of possible electronic transitions. For example, transitions between the lower band of the $S$ cone and the upper band of the $F$ cone, represented as $S_-\leftrightarrow F_+$ (or equivalently as $|\Psi_{-,+}\rangle \leftrightarrow |\Psi_{+,+}\rangle$), are accounted for by the two terms in Eq. (\ref{Eq:Trace_definition}) that contain both vectors $|\Psi_{-,+}\rangle$ and $|\Psi_{+,+}\rangle$. We  denote those terms as 

\begin{figure}[h]
\includegraphics[width=0.44\textwidth]{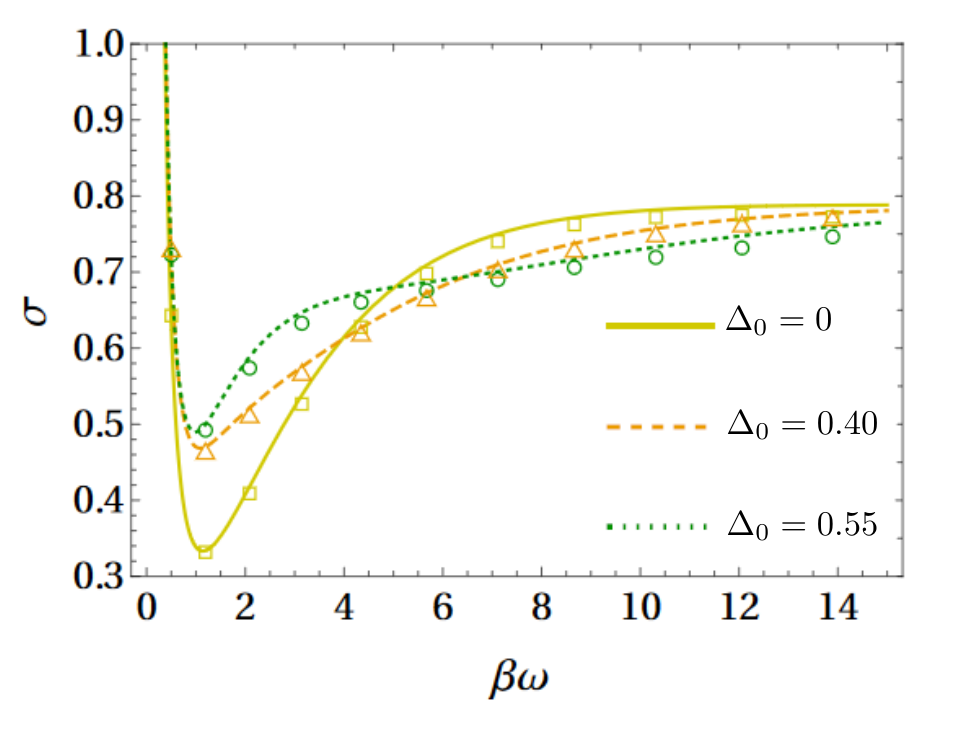}
\caption{\label{Fig:Sigma_total} Total electronic conductivity $\sigma$ (in units of $2e^2/h$) of kekul\'e-patterned graphene for different values of the coupling amplitude $\Delta_0$ and a scattering rate of  $\beta\eta=5\times 10^{-2}$. The curves are given by the analytical expressions in Eqs. (\ref{Eq.Sigma_inter}-\ref{Eq.Sigma_Y}) and the symbols show the respective numerical calculations obtained from Eq. (\ref{Eq:Sigma_2}).}
\end{figure}

\begin{eqnarray}\label{Eq:Trace_A_def}
    T_{S_-\leftrightarrow F_+}
    &&=\int\{\langle \Psi_{ -,+}|\Lambda_+|\Psi_{+,+} \rangle\langle \Psi_{+,+}| \Lambda_-|\Psi_{-,+}\rangle \\ && +\langle \Psi_{ +,+}|\Lambda_+|\Psi_{-,+} \rangle\langle \Psi_{-,+}| \Lambda_-|\Psi_{+,+}\rangle \} \frac{d^2\xi}{(2\pi)^2} \nonumber
\end{eqnarray}

and do similarly for all of the other terms in Eq. (\ref{Eq:Trace_expanded}). 
To give a clear physical picture of all these kind of processes, in Figs. \ref{Fig:Intervalley_transitions} and \ref{Fig:Intravalley_transitions} we show a sketch of all the different types of intervalley and intravalley transitions. 
Next we solve Eq. (\ref{Eq:Trace_A_def}) by using polar coordinates $d^2\xi \rightarrow \xi d\xi d\theta$ and defining the energy cutoff by $\lambda$. After the angular integration one has

\begin{eqnarray}\label{Eq:Trace_A_1}
    T_{S_-\leftrightarrow F_+}(\epsilon)=&&\frac{\pi}{(2 \pi)^2} \int_0^\lambda [\delta_\eta(E_{+,+}-\epsilon_+)\delta_\eta(E_{-,+}-\epsilon_-) \nonumber \\ 
    && \delta_\eta(E_{-,+}-\epsilon_+)\delta_\eta(E_{+,+}-\epsilon_-)]\xi d\xi
\end{eqnarray}
 
Considering that $\eta$ and $\Delta_0$ are both small parameters, we further approximate   $\eta\Delta_0 \rightarrow 0$ so we can write, after substituting the energy eigenvalues,
\begin{eqnarray}\label{Eq:Trace_2}
    T_{S_-\leftrightarrow F_+}&&(\epsilon)=\frac{\pi}{(2 \pi)^2}
    \frac{1}{1-\Delta_0^2}\times \nonumber \\  && \int_0^\lambda \Bigl[\delta_\eta\Bigl(\xi-\frac{\epsilon_+}{1+\Delta_0}\Bigr) \delta_\eta\Bigl(\xi+\frac{\epsilon_-}{1-\Delta_0}\Bigr) \nonumber \\ 
    &&\delta_\eta\Bigl(\xi+\frac{\epsilon_+}{1-\Delta_0}\Bigr) \delta_\eta\Bigl(\xi-\frac{\epsilon_-}{1+\Delta_0}\Bigr)\Bigr] \xi d\xi
\end{eqnarray}

Treating the trace terms for all types of transitions in a similar way and adding the results gives (see Appendix)

\begin{eqnarray}\label{Eq:Trace_solved}
    T(\epsilon)&&=\frac{\pi}{(2\pi)^2}\Biggl\{\frac{1}{2}(\omega/2 + \Delta_0\epsilon) \delta_\eta(\epsilon+ \Delta_0\omega /2)\frac{1}{1-\Delta_0^2}
    \nonumber \\ &&+\frac{1}{2}(\omega/2-\Delta_0\epsilon) \delta_\eta(\epsilon-\Delta_0\omega  /2)\frac{1}{1-\Delta_0^2} \nonumber \\ &&
    +\frac{\epsilon \delta t}{\pi}\delta_\eta(\omega/2)+\frac{\delta t}{2\pi}\frac{\Delta_0^2}{1-\Delta_0^2}[(\epsilon+\Delta_0\omega/2)\delta_\eta(\Delta_0\epsilon+\omega/2) \nonumber \\ && 
    +(\epsilon-\Delta_0\omega/2)\delta_\eta(\Delta_0 \epsilon-\omega/2)]
     \nonumber \\ 
     &&
     +\frac{2\eta}{\pi\omega}\biggl[1+\frac{\Delta_0^2}{1 - (2\Delta_0\epsilon/\omega)^2}
      \biggr]\Biggr\}\Theta(\lambda-\omega/2)
\end{eqnarray}

where $\delta t =[\atan(\epsilon_+/\eta)+\atan(\epsilon_-/\eta)]$. The first two terms describe $S_- \leftrightarrow F_+$ and $S_+\leftrightarrow F_-$ transitions,  which are interband transitions ($E\rightarrow E\pm \omega$) as shown in Fig. \ref{Fig:Intervalley_transitions}. These become the usual optical transitions in graphene when $\Delta_0\rightarrow0$, however, in Kek-Y graphene ($\Delta_0>0$), this transitions involve a change in velocity (or cone). The third term in Eq. (\ref{Eq:Trace_solved}) describes all of the intraband transitions shown in Fig. \ref{Fig:Intravalley_transitions}.
The fourth term describes the vertical transitions $S_+ \leftrightarrow F_+$ and $S_-\leftrightarrow F_-$ also shown in Fig. \ref{Fig:Intervalley_transitions}; these are intervalley transitions that are absent in pristine graphene, as can be seen by taking $\Delta_0 \rightarrow 0$. These transitions, like interband transitions, imply a change in energy $E\rightarrow E\pm\omega$, but happen between states in the conduction band or between states in the valence band, like intraband transitions. Interestingly, the terms related to the intravalley transitions $S_-\leftrightarrow S_+$ and $F_-\leftrightarrow F_+$ shown in Fig. \ref{Fig:Intravalley_transitions} contain the factors $\langle \Psi_{+,+}|\partial H /\partial \xi_y |\Psi_{-,-}\rangle$ and $\langle \Psi_{-,+}|\partial H /\partial \xi_y |\Psi_{+,-}\rangle$ which according to Eq. (\ref{Eq:Current_op_diag}) are zero, and therefore, only interband transitions which change the velocity are allowed for $\Delta_0>0$. In section \ref{Absence} we will use Fermi's golden rule to further verify this result.
The last term in Eq. (\ref{Eq:Trace_solved}) is related to horizontal transitions $S_\pm \leftrightarrow F_\pm$ and $S_\mp\leftrightarrow F_\pm$ shown in Figs. \ref{Fig:Intervalley_transitions} and \ref{Fig:Intravalley_transitions}. This term is of higher order in $\eta$ and its contribution to the dynamical conductivity is negligible. However, in some of the minimal conductivities it adds a term of the order $\Delta_0^2e^2/h$ (see \ref{Sec.Minimal_conds}).

We first obtain the interband conductivity  $\sigma_{inter}$ by assuming $\omega < 2\lambda$ and substituting the first two terms of $T(\epsilon)$ in Eq. (\ref{Eq:Sigma_2}) and multiplying by a factor of 2 in order to take spin degeneracy into account. This leads to

\begin{equation}\label{Eq.Sigma_inter}
    \sigma_{inter} = \frac{\pi e^2}{2 h}\frac{\sinh(\beta \omega/2)}{\cosh(\beta \omega /2)+ \cosh(\Delta_0\beta \omega/2)}
\end{equation}

As expected, for the case of $\Delta_0=0$ this expression reduces to the interband conductivity of pristine graphene found in Refs. \cite{Mak,Ziegler,Ryu,Koshino2008,Gusynin}. 

On the other hand, the intraband conductivity $\sigma_{intra}$ is obtained by substituting the third term of Eq. (\ref{Eq:Trace_solved}) into Eq. (\ref{Eq:Sigma_2}) and similarly taking spin degeneracy into account. Assuming $\omega <2\lambda$,

\begin{equation}\label{Eq:Sigma_Intra}
    \sigma_{intra}=4\ln2\frac{e^2}{h}\frac{2\beta\eta}{(\beta\omega)^2 + (2\beta\eta)^2}.
\end{equation}

Lastly, the  fourth term in $T(\epsilon)$ gives
\begin{equation}\label{Eq.Sigma_Y}
    \sigma_Y=\frac{\pi 
    e^2}{2h}\frac{\sinh(\beta\omega/2)}{\cosh(\beta\omega/2)+\cosh(\Delta_0^{-1}\beta\omega/2)} 
\end{equation}
which is the component of the conductivity that describes vertical (or optical) intervalley transitions $S_+ \leftrightarrow F_+$ and $S_-\leftrightarrow F_-$ (see Fig. \ref{Fig:Intervalley_transitions}).

\section{Discussion}\label{Sec:Discussion}

This section is divided in several subsections in which we discuss each kind of  channel, including plots of each contribution to the conductivity. We will also include a study of the diverse
physical limits imposed to the conductivity.

Before going into such details, let us study first  the total conductivity $\sigma=\sigma_{inter}+\sigma_{intra}+\sigma_Y$. At this point, it is worthwhile mentioning that in order to test the accuracy of the approximations used to obtain the analytical expressions in Eqs. (\ref{Eq.Sigma_inter}-\ref{Eq.Sigma_Y}), we also performed a direct and independent numerical evaluation. 
Therefore, in Fig. \ref{Fig:Sigma_total}  we present the typical behavior of $\sigma$ for different values of $\Delta_0$, in this case by using a constant scattering rate of $\beta \eta=5 \times 10^{-2}$. The symbols  in Fig. \ref{Fig:Sigma_total} were obtained from a direct numerical calculation using Eqs. (\ref{Eq:Sigma_2}) and (\ref{Eq:Trace_expanded}) and a mesh in 
$\boldsymbol{k}$-space. The curves were obtained using the analytical expressions for different values of the coupling amplitude $\Delta_0$.

An excellent agreement between the analytical and numerical results for values of  $\beta\eta < 10^{-1}$ is obtained. At room temperature, this corresponds to scattering rates $\eta$ with values up to tens of meV (in units of $\hbar$), which is the typical range reported for graphene samples \cite{Mak}. For higher values of $\beta\eta$, we find that $\sigma_{inter}$ starts to decrease upon increasing scattering rate. 

We can distinguish in Fig. \ref{Fig:Sigma_total} a general overall behavior. For a fixed temperature and at a high frequency, the almost flat-frequency response behaviour of pristine   graphene is recovered. However, the  conductivity is reduced by the kekul\'e ordering as it introduces scattering. As expected for the low-frequency region, a universal Drude-peak due to disorder scattering is observed. In the crossover region between both limits, the conductivity's shape depends a lot upon disorder, temperature, frequency, chemical potential and kekul\'e ordering. This requires a careful understanding of each contribution as detailed in the following subsections. Before going further, we point out that the low-energy dispersion used here [Eq. (\ref{Eq:Dispersion})] remains valid for high values of the coupling parameter $\Delta_0$ upon a simple renormalization of the velocities $v_0(1\pm\Delta_0)$. Here, this renormalization can be done straightforwardly by substituting $\Delta_0 \rightarrow \Delta_0'=\rho_-/\rho_+$ in all of our results, with $\rho_-$ and $\rho_+$ defined as in Ref. \cite{Gamayun}.

\subsection{Interband conductivity}\label{Subsec.Interband}

\begin{figure}[t]
\includegraphics[width=0.45\textwidth]{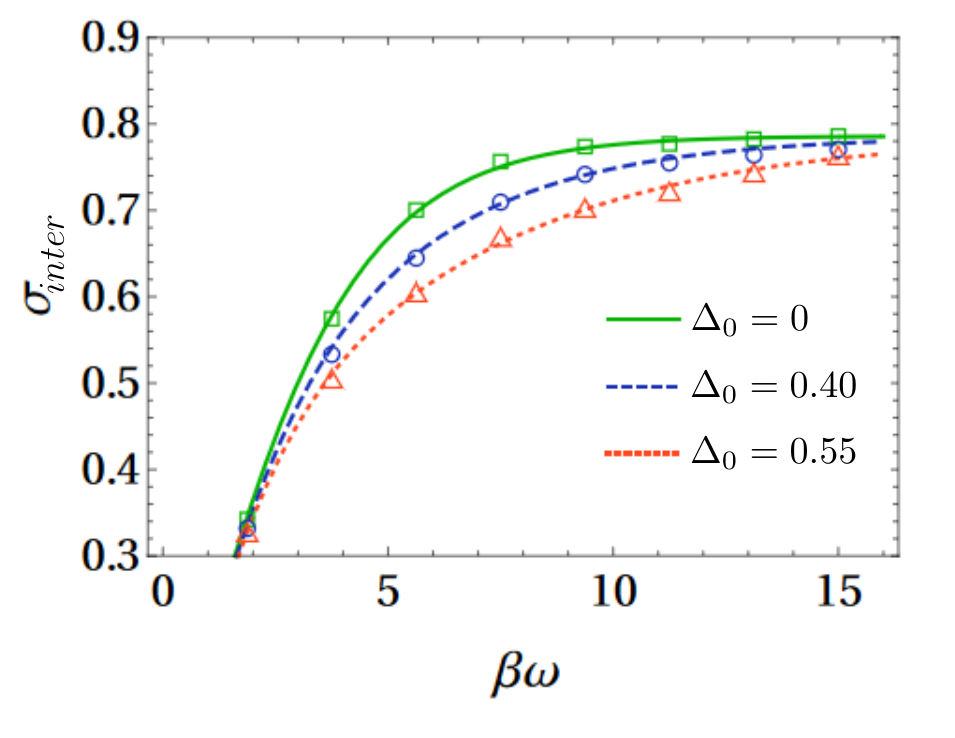}
\caption{\label{Fig:Sigma_inter} Interband conductivity (in units of $2e^2/h$) for different values of the coupling amplitude $\Delta_0$ and a constant scattering rate $\beta\eta=10^{-2}$. The curves are given by the analytical expression in Eq. (\ref{Eq.Sigma_inter}) and the symbols show the respective numerical calculations obtained directly from Eq. (\ref{Eq:Sigma_2}). The values  for $\Delta_0$ were chosen so to have a better visualization. Realistic values are expected to be lower than the ones shown here.}
\end{figure}

\begin{figure*}
\includegraphics[width=0.9\textwidth]{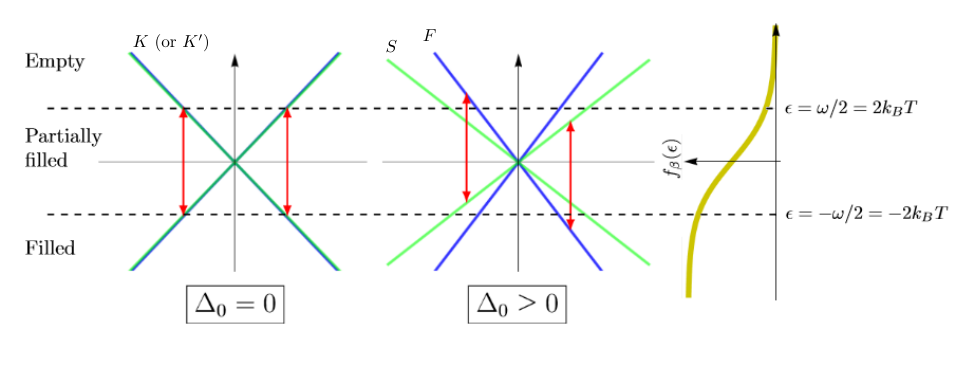}
\caption{Diagram of interband transitions around $\beta\omega/4\approx1$. Interband transitions in non-distorted graphene ($\Delta_0=0$) occur just in the limits of the energetic region of partially filled states (defined as $\epsilon$ such that $0.1\lesssim f_\beta(\epsilon)\lesssim0.9$), while the interband transitions in kekul\'e-distorted graphene ($\Delta_0>0$) involve states in the partially filled region, where there are less states available for transitions and therefore $\sigma_{inter}$ decreases with $\Delta_0$ around intermediate values of $\beta\omega$.
 \label{Fig:Thermal_effect}}.
\end{figure*}

Plots of $\sigma_{inter}$ are shown in Fig. \ref{Fig:Sigma_inter} for different values of $\Delta_0$. In this figure we compare the analytical expressions with direct numerical evaluations of Eqs. (\ref{Eq:Sigma_2}) and (\ref{Eq:Trace_expanded}). All the conductivities are plotted in units of $g_s e^2/ h$, with $g_s=2$ the spin degeneracy. A decrease in $\sigma_{inter}$ is found with increasing $\Delta_0$. A sketch explaining this behavior is shown in Fig. \ref{Fig:Thermal_effect}. In pristine graphene ($\Delta_0=0$), for values of $\beta\omega/4\lesssim 1$ the states that can participate in interband transitions (those satisfying $E_f(\boldsymbol{k})-E_i(\boldsymbol{k})=\omega$) lie in the partially-filled energy region, which can be roughly defined by $0.1\lesssim f_\beta(\epsilon)\lesssim 0.9$, where less states are available for transitions. Therefore, the interband conductivity drops to zero as $\beta\omega$ decreases (the factor $f_\beta(\epsilon_+)-f_\beta(\epsilon_-)$ in Eq. (\ref{Eq:Sigma_2}) takes lower values). In kekul\'e-patterned graphene ($\Delta_0 >0$), the energy of the states that can participate in interband transitions shifts, leading this effect to occur in a wider range of $\beta\omega$. This reduction of $\sigma_{inter}$ due to the interplay between temperature and an energy shift in the interband transitions is similar to that  reported recently for anisotropic tilted Dirac semimetals \cite{Borophene}.

\subsection{Intraband conductivity}\label{Subsec.Sigma_intra}

The intraband conductivity $\sigma_{intra}$ is the Drude peak known to be present in graphene's conductivity for low photon frequencies \cite{Mak,Falkovsky,Kuzmenko,Horng,Li2008} where intraband transitions become important. 

\begin{figure}[t]
\includegraphics[width=0.45\textwidth]{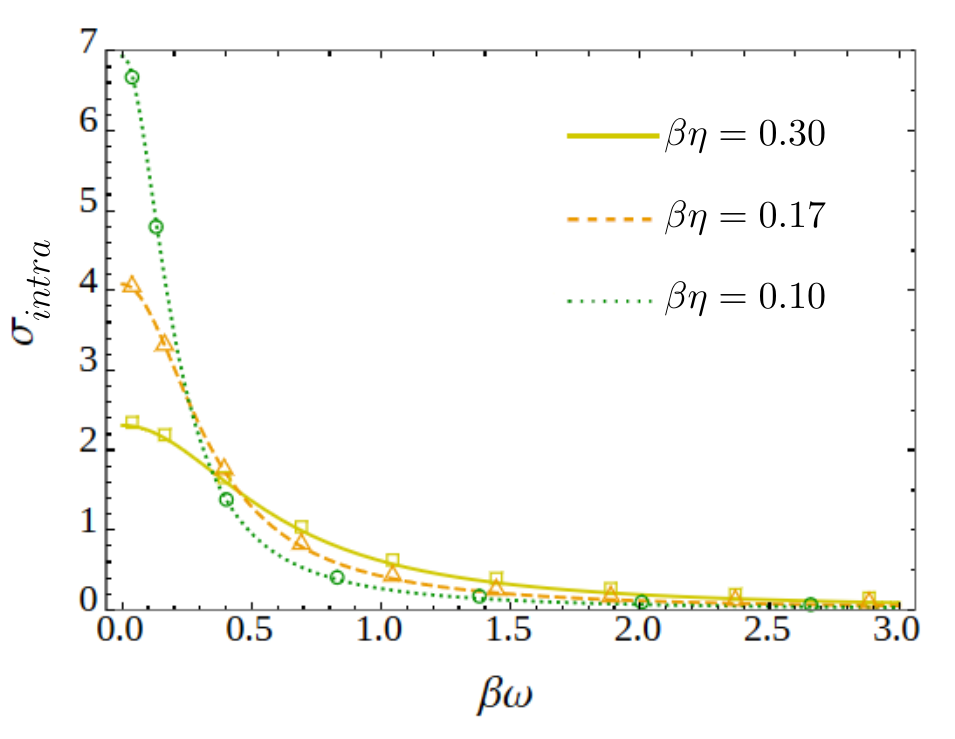}
\caption{\label{Fig:Sigma_intra_graphene} Intraband conductivity (in units of $2e^2/h$) for different values of the scattering rate $\beta\eta$. The curves are given by the analytical expression in Eq. (\ref{Eq:Sigma_Intra}) and the symbols show the respective numerical calculations obtained from Eq. (\ref{Eq:Sigma_2}).}
\end{figure}

Our results for $\sigma_{intra}$ reproduce the Drude contribution used to fit experimental measurements in Ref. \cite{Mak} with a phenomenological scattering rate given by $\Gamma=2 \eta$. This important feature was not obtained in a similar treatment of the  Kubo conductivity for graphene in Ref. \cite{Ziegler}, where the same Eqs. (\ref{Eq:Sigma_2}) and (\ref{Eq:Trace_expanded}) were solved. The reason is that that they were interested in the high-frequency region and thus different approximations were made. Therefore, the low-frequency conductivity did not reproduce the Lorentzian behavior. In Fig.  \ref{Fig:Sigma_intra_graphene} we plot $\sigma_{intra}$ comparing the analytical expression in Eq. (\ref{Eq:Sigma_Intra}) with the numerical evaluation for different values of $\beta\eta$.

\subsection{Intervalley conductivity $\sigma_Y$}\label{Subsec.Sigma_Y}
\begin{figure}[h]
\includegraphics[width=0.48\textwidth]{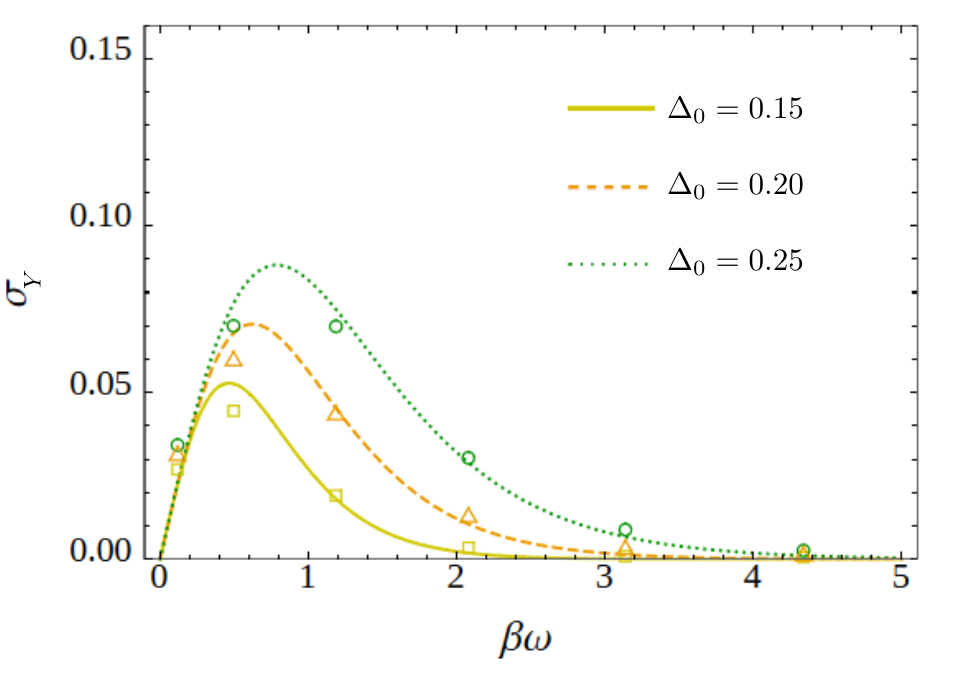}\caption{Intervalley conductivity $\sigma_Y$ (in units of $2e^2/h$) for different values of $\Delta_0$ and a constant scattering rate of $\beta\eta=5\times10^{-2}$. This component arises as a consequence of the intervalley currents introduced by $\Delta_0>0$. The curves are given by the analytical expression in Eq. (\ref{Eq.Sigma_Y}) and the symbols show the respective numerical calculations.
 \label{Fig.SigmaY}}
\end{figure}

This component is plotted in Fig. \ref{Fig.SigmaY}. It is zero for the case of pristine graphene and grows in amplitude as $\Delta_0$ increases. For small values of $\Delta_0$ and $\beta\eta$, it displays a maximum around $\beta\omega\approx \pi\Delta_0$. The dependence of $\sigma_Y$ as a function of $\beta\omega$ can be understood with arguments similar to those used for $\sigma_{inter}$. The transitions in this case, however, involve both the upper bands ($S_+ \leftrightarrow F_+$) or the lower bands ($S_- \leftrightarrow F_-$). Furthermore, the introduction of a chemical potential $\beta \mu\neq 0$ can shift this transitions in frequency and amplitude, as we discuss below.

\subsection{Tuning $\sigma_Y$ with a chemical potential   $\mu$}

\begin{figure}[t]
\includegraphics[width=0.45\textwidth]{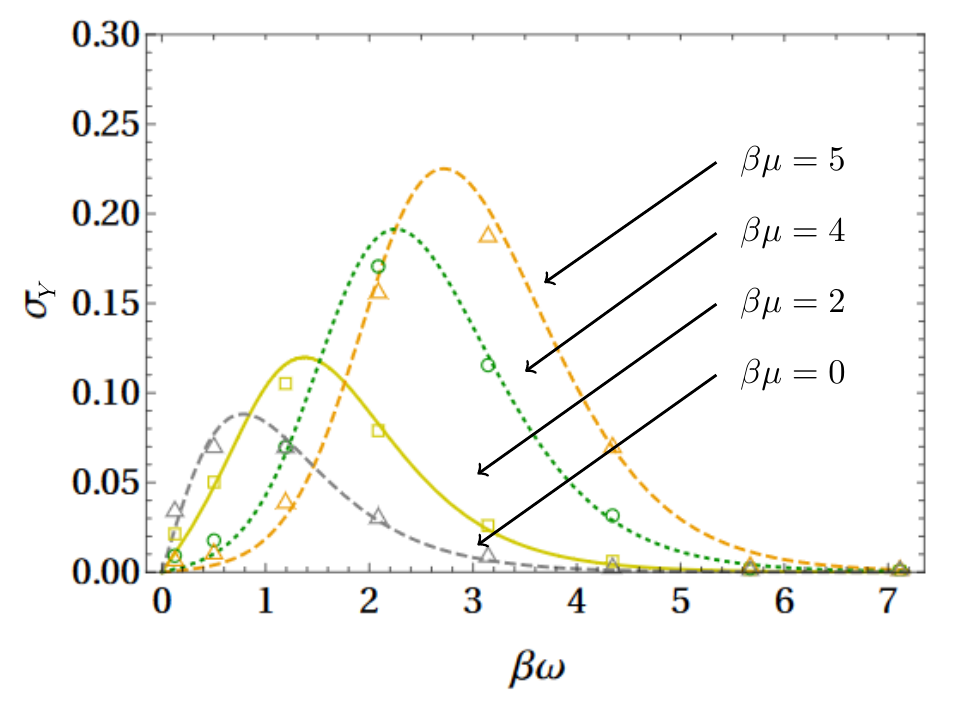}
\caption{\label{Fig.SigmaY_ChemPot}Intervalley conductivity $\sigma_Y$ (in units of $2e^2/h$) for different values of the chemical potential $\beta\mu$ and a constant $\beta\eta=5\times10^{-2}$ and $\Delta_0=0.25$. The absorption maximum due to intervalley transitions can be shifted by varying doping. The curves are given by the analytical expression in Eq. (\ref{Eq.Sigma_Y_mu}) and the symbols show the respective numerical calculations.}
\end{figure}

To take into account the effect of a finite doping, a chemical potential can be easily introduced through the Fermi-Dirac distributions $f_\beta$ in Eq.  (\ref{Eq:Sigma_2}). This is an important factor, as graphene samples under normal conditions tend to present spontaneous doping \cite{Mak}.
The generalization of each component of the conductivity for a finite $\mu$ results in,

\begin{eqnarray}\label{Eq.Sigma_inter_mu}
    \sigma_{inter}=&&\frac{\pi e^2}{4 h}\biggl[ \frac{\sinh(\beta\omega/2)}{\cosh(\beta\omega/2)+\cosh(\Delta_0\beta\omega/2+\beta\mu)} \nonumber \\ &&
    + \frac{\sinh(\beta\omega/2)}{\cosh(\beta\omega/2)+\cosh(\Delta_0\beta\omega/2-\beta\mu)}\biggr],
\end{eqnarray}

\begin{eqnarray}\label{Eq.Sigma_Y_mu}
    \sigma_{Y}=&&\frac{\pi e^2}{4 h}\biggl[ \frac{\sinh(\beta\omega/2)}{\cosh(\beta\omega/2)+\cosh(\Delta_0^{-1}\beta\omega/2+\beta\mu)} \nonumber \\ &&
    + \frac{\sinh(\beta\omega/2)}{\cosh(\beta\omega/2)+\cosh(\Delta_0^{-1}\beta\omega/2-\beta\mu)}\biggr],
\end{eqnarray}

\begin{equation}\label{Eq.Sigma_intra_mu}
    \sigma_{intra}=\frac{2e^2}{h}\ln[2+2\cosh(\beta\mu)]\frac{2\beta\eta}{(\beta\omega)^2+(2\beta\eta)^2}
\end{equation}
As expected, the effect of having $\beta\mu\neq0$ in $\sigma_{inter}$ is essentially to block transitions for $\beta\omega<2\beta\mu$, while leaving the behavior for $\beta\omega >2\beta\mu$ practically unchanged. For $\sigma_Y$, the effect of a finite chemical potential is more interesting. A plot of $\sigma_Y$ for different values of $\beta\mu$ is shown in Fig. \ref{Fig.SigmaY_ChemPot}. It can be seen that the amplitude, as well as the position of the maximum increase with $\beta\mu$. The maximum is located around $\beta\omega\approx\pi\Delta_0+\pi\Delta_0\beta\mu/2$. As we show in Fig. \ref{Fig:Chemical_Intervalley_transitions},
states far below $\beta\mu$ are all fully occupied and states far above $\beta\mu$ are all empty, therefore, the energy of the states that can participate in $S_+\leftrightarrow F_+$ (or $S_-\leftrightarrow F_-$) transitions moves together with $\beta\mu$. A larger value of $\beta\mu$ implies these transitions occurring at higher energies, where the density of states is larger (which leads to larger amplitudes of $\sigma_Y$) and where there is a larger separation between the cones (which leads to the transitions occurring at higher frequencies). Lastly, $\sigma_{intra}$ also increases in amplitude with $\beta\mu$ because the states available for scattering are located at higher energies (around $\beta\mu$), where the density of states is larger. For a high enough chemical potential, the maximum of $\sigma_Y$ will be located in the region where $\sigma_{inter}$ is zero due to Pauli blocking, making it possible to measure $\sigma_Y$ alone in a wide range of frequencies.

\subsection{Absence of $S_-\leftrightarrow S_+$ and $F_-\leftrightarrow F_+$ transitions}\label{Absence}

The result that interband transitions within a valley ($S_-\leftrightarrow S_+$ and $F_-\leftrightarrow F_+$ transitions) are absent for $\Delta_0\neq0$ can be further verified by using Fermi's golden rule. First, the electric field $\boldsymbol{E}$ is introduced as a perturbation $\delta H$ to the Hamiltonian in Eq. (\ref{Eq:Hamiltonian}) through the minimal coupling:
\begin{figure}[t]
\includegraphics[width=0.48\textwidth]{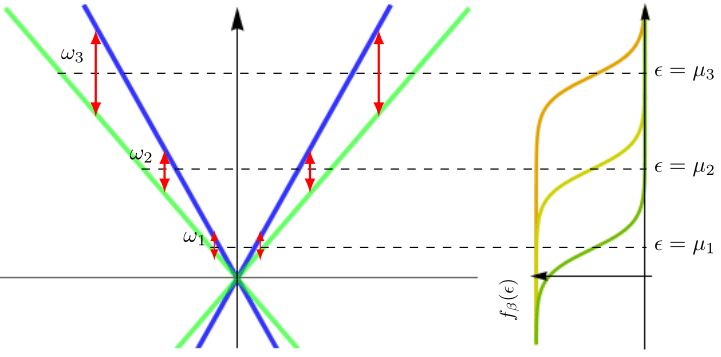}
\caption{\label{Fig:Chemical_Intervalley_transitions} Diagram of transitions between the upper bands that contribute to the intervalley conductivity $\sigma_Y$ for different chemical potentials. At a constant temperature, for larger chemical potentials ($\beta\mu_3>\beta\mu_2>\beta\mu_1$), the transitions occur at higher frequencies ($\beta\omega_3>\beta\omega_2>\beta\omega_1$), resulting in $\sigma_Y$ having the behavior shown in Fig. \ref{Fig.SigmaY_ChemPot}. Similar transitions can occur between the lower bands (not shown).}

\end{figure}
\begin{equation}\label{Eq:Hamiltonian_perturbation}
    H=(\boldsymbol{p}-\frac{e}{c}\boldsymbol{A})\cdot\boldsymbol{\sigma'}\otimes\tau_0 + \Delta_0\sigma_0\otimes(\boldsymbol{p}-\frac{e}{c}\boldsymbol{A})\cdot\boldsymbol{\tau'} \equiv H_0+\delta H 
\end{equation}

where $\boldsymbol{\sigma'}=v_0\boldsymbol{\sigma}/\hbar$, $\boldsymbol{\tau'}=v_0 \boldsymbol{\tau}/\hbar$, $\boldsymbol{p}=\hbar \boldsymbol{k}$ and $\partial \boldsymbol{A}/\partial t = -c\boldsymbol{E}$. Therefore the perturbation is given by \cite{KATSNELSON200720}

\begin{equation}\label{Eq:Perturbation}
    \delta H=\frac{i e}{2 \omega}\boldsymbol{E}\cdot\boldsymbol{\sigma'}\otimes \tau_0 + \Delta_0 \frac{i e}{2 \omega}\sigma_0\otimes \boldsymbol{E}\cdot\boldsymbol{\tau'}
\end{equation}

Then, according to Fermi's golden rule, the transition probability is proportional to $\langle \Psi_{+,-}|\delta H | \Psi_{-,+} \rangle$ for the $S_-\leftrightarrow S_+$ transitions and to $\langle \Psi_{+,+}|\delta H | \Psi_{-,-} \rangle$ for the $F_-\leftrightarrow F_+$ transitions. Using the orthogonality of the eigenvector basis it can be verified that these amplitudes are equal to zero.
This can be understood by considering the chirality of the $S$ and $F$ cones. When $\Delta_0 >0$, the $S$ and $F$ cones inherit the chirality of the $K$ and $K'$ valleys in non-distorted graphene, in such a way that the $S$ cone is completely chiral and the $F$ cone is completely anti-chiral. The transitions $S_-\leftrightarrow S_+$ and $F_-\leftrightarrow F_+$ transitions are thus forbidden.

\subsection{Minimal conductivities}\label{Sec.Minimal_conds}

As is well known, in graphene and other Dirac materials the Kubo formula leads to multiple minimal conductivities when temperature, frequency and scattering limits are taken in different orders \cite{Ziegler,Borophene,Mak}.
We calculate first the zero temperature limits. When $\beta \rightarrow \infty$, Eq. (\ref{Eq:Sigma_2}) reduces to the integral of $T(\epsilon)/\omega$ from $\epsilon=-\omega/2$ to $\epsilon=\omega/2$. Substituting Eq. (\ref{Eq:Trace_solved}) and taking into account spin degeneracy results in

\begin{equation}\label{Eq:MinCondZeroT}
\sigma\approx \frac{\pi}{2}\Bigl[1+(1+\Delta_0^2)\frac{4\eta}{\pi\omega}\Bigr]\frac{e^2}{h}
\end{equation}
which leads to 

\begin{equation}
    \sigma^{min}_1\approx\frac{\pi}{2}\frac{e^2}{h}, \qquad \text{for $\omega\gg\eta$,
    }
\end{equation}

\begin{equation}
    \sigma^{min}_2\approx\pi\frac{e^2}{h}\Bigl[1+\frac{1}{2}\Delta_0^2\Bigr], \qquad \text{for $\eta\approx\omega$}.
\end{equation}
One more value is obtained if  we first take the limit $\omega\rightarrow 0$ in Eq. (\ref{Eq:Sigma_2}). This gives $\sigma_3^{min}\approx \pi e^2T(0)/\hbar$ which leads to 
\begin{equation}
    \sigma^{min}_3 \approx 8\pi\frac{e^2}{h}(1+\Delta_0^2)\int_0^\lambda\frac{(\eta/\pi)^2}{(\xi^2+\eta^2)^2}\xi d \xi=\frac{4}{\pi}\frac{e^2}{h}(1+\Delta_0^2) 
\end{equation}

If we now consider a temperature and frequency dependence,  for the expressions in Eqs. (\ref{Eq.Sigma_inter}-\ref{Eq.Sigma_Y}) we get 

\begin{equation}
    \sigma \approx \frac{2e^2}{h}\Bigg\{\begin{array}{cc}
        \text{$ \;\,\ln2 /\beta\eta$ \qquad \qquad for $\beta\omega\sim 0$}, \\
         \text{$\pi/4 +4\ln2 \beta\eta/(\beta\omega)^2$ \quad for $\beta\omega\sim\infty$}.
    \end{array}
\end{equation}

These last expressions do not depend on the coupling amplitude $\Delta_0$.

\section{Conclusions}
We have obtained analytical expressions for the electronic and optical conductivities of the recently seen kekulé-patterned graphene superlattices \cite{Gutierrez2016}, for which a momentum-valley coupling was predicted \cite{Gamayun}. Our results take into consideration the dependence on frequency, temperature, disorder and chemical potential. When compared with pristine graphene, new terms to the conductivity were found, as a result of the opening of intervalley channels that are not present in non-distorted graphene. The transitions involved are predicted to be tunable in frequency and amplitude by the chemical potential. Direct numerical calculations were also performed and compared. Our analytical results correctly reduce to that of graphene in the appropriate limits.
The results show that for a fixed temperature and at low-frequencies, the conductivity presents the same Drude peak as in graphene while in the high-frequency limit, there is an asymptote towards the flat behavior, as also seen in graphene. The cross-over region presents an interesting interplay between the kekul\'e ordering, disorder, temperature and chemical potential. Finally, we also provided the minimal conductivities in the different physical limits concerning such parameters. 

\begin{acknowledgments}
We thank UNAM DGAPA-PROJECT IN102620. S. A. H. thanks CONACYT for a MSc scholarship. We also thank Elias Andrade and Ramon Carrillo-Bastos for critical comments and discussions
on this work.  
\end{acknowledgments}

\appendix

\section{\label{Appendix} Double delta integral}

The complete expression for the double delta integral in the expression for $T(\epsilon)$ is

\begin{eqnarray}\label{Eq.Double_delta_full}
\int_0^\lambda && \delta_\eta(\xi-a)\delta_\eta(\xi-b)\xi d\xi = \frac{(a+b)}{8\pi}\delta_\eta\bigl(\frac{a-b}{2}\bigr) \\ \nonumber &&\times[\atan(a/\eta)+\atan(b/\eta)+\atan\bigl(\frac{\lambda-a}{\eta}\bigr) \\ \nonumber &&+\atan\bigl(\frac{\lambda-b}{\eta}\bigr)]-\frac{\eta}{2\pi^2 (a-b)}[\atan\bigl(\frac{\lambda-b} {\eta}\bigr) \\ \nonumber && +\atan(b/\eta)-\atan\bigl(\frac{\lambda-a}{\eta}\bigr)-\atan(a/\eta)] \\ \nonumber && +\frac{\eta^2}{2\pi^2}\frac{(a+b)/(a-b)}{[(a-b)^2+4\eta^2]}\ln\left[\frac{a^2+\eta^2}{b^2+\eta^2}\times\frac{(\lambda-a)^2+\eta^2}{(\lambda-b)^2+\eta^2}\right]
\end{eqnarray}

assuming $\lambda>\eta$ and that $a$ and $b$ are not simultaneously negative (otherwise we take the integral equal to zero).  
For a small $\eta$ and $\lambda\gg a,b,\eta$, this integral can be approximated by

\begin{eqnarray}\label{Eq.Double_delta_approx}
\int_0^\lambda && \delta_\eta(\xi-a)\delta_\eta(\xi-b)\xi d\xi \approx \frac{(a+b)}{8\pi}\delta_\eta\bigl(\frac{a-b}{2}\bigr) \\ \nonumber &&\times[\atan(a/\eta)+\atan(b/\eta)+\pi].
\end{eqnarray}
To obtain the last term in Eq. (\ref{Eq:Trace_solved}), which can be neglected in the dynamical conductivity but produces a term of order $\Delta_0^2 e^2/h$ in some of the minimal conductivities, one needs to take into account the second term in Eq. (\ref{Eq.Double_delta_full}). All of the remaining terms in Eq. (\ref{Eq:Trace_solved}) can be obtained by this approximation, however.

\nocite{*}

\bibliography{References_KekY}

\providecommand{\noopsort}[1]{}\providecommand{\singleletter}[1]{#1}%
\begin{thebibliography}{32}%
\makeatletter
\providecommand \@ifxundefined [1]{%
 \@ifx{#1\undefined}
}%
\providecommand \@ifnum [1]{%
 \ifnum #1\expandafter \@firstoftwo
 \else \expandafter \@secondoftwo
 \fi
}%
\providecommand \@ifx [1]{%
 \ifx #1\expandafter \@firstoftwo
 \else \expandafter \@secondoftwo
 \fi
}%
\providecommand \natexlab [1]{#1}%
\providecommand \enquote  [1]{``#1''}%
\providecommand \bibnamefont  [1]{#1}%
\providecommand \bibfnamefont [1]{#1}%
\providecommand \citenamefont [1]{#1}%
\providecommand \href@noop [0]{\@secondoftwo}%
\providecommand \href [0]{\begingroup \@sanitize@url \@href}%
\providecommand \@href[1]{\@@startlink{#1}\@@href}%
\providecommand \@@href[1]{\endgroup#1\@@endlink}%
\providecommand \@sanitize@url [0]{\catcode `\\12\catcode `\$12\catcode
  `\&12\catcode `\#12\catcode `\^12\catcode `\_12\catcode `\%12\relax}%
\providecommand \@@startlink[1]{}%
\providecommand \@@endlink[0]{}%
\providecommand \url  [0]{\begingroup\@sanitize@url \@url }%
\providecommand \@url [1]{\endgroup\@href {#1}{\urlprefix }}%
\providecommand \urlprefix  [0]{URL }%
\providecommand \Eprint [0]{\href }%
\providecommand \doibase [0]{https://doi.org/}%
\providecommand \selectlanguage [0]{\@gobble}%
\providecommand \bibinfo  [0]{\@secondoftwo}%
\providecommand \bibfield  [0]{\@secondoftwo}%
\providecommand \translation [1]{[#1]}%
\providecommand \BibitemOpen [0]{}%
\providecommand \bibitemStop [0]{}%
\providecommand \bibitemNoStop [0]{.\EOS\space}%
\providecommand \EOS [0]{\spacefactor3000\relax}%
\providecommand \BibitemShut  [1]{\csname bibitem#1\endcsname}%
\let\auto@bib@innerbib\@empty
\bibitem [{\citenamefont {Guti{\'e}rrez}\ \emph {et~al.}(2016)\citenamefont
  {Guti{\'e}rrez}, \citenamefont {Kim}, \citenamefont {Brown}, \citenamefont
  {Schiros}, \citenamefont {Nordlund}, \citenamefont {Lochocki}, \citenamefont
  {Shen}, \citenamefont {Park},\ and\ \citenamefont
  {Pasupathy}}]{Gutierrez2016}%
  \BibitemOpen
  \bibfield  {author} {\bibinfo {author} {\bibfnamefont {C.}~\bibnamefont
  {Guti{\'e}rrez}}, \bibinfo {author} {\bibfnamefont {C.-J.}\ \bibnamefont
  {Kim}}, \bibinfo {author} {\bibfnamefont {L.}~\bibnamefont {Brown}}, \bibinfo
  {author} {\bibfnamefont {T.}~\bibnamefont {Schiros}}, \bibinfo {author}
  {\bibfnamefont {D.}~\bibnamefont {Nordlund}}, \bibinfo {author}
  {\bibfnamefont {E.}~\bibnamefont {Lochocki}}, \bibinfo {author}
  {\bibfnamefont {K.~M.}\ \bibnamefont {Shen}}, \bibinfo {author}
  {\bibfnamefont {J.}~\bibnamefont {Park}},\ and\ \bibinfo {author}
  {\bibfnamefont {A.~N.}\ \bibnamefont {Pasupathy}},\ }\bibfield  {title}
  {\bibinfo {title} {Imaging chiral symmetry breaking from kekul{\'e} bond
  order in graphene},\ }\href {https://doi.org/10.1038/nphys3776} {\bibfield
  {journal} {\bibinfo  {journal} {Nature Physics}\ }\textbf {\bibinfo {volume}
  {12}},\ \bibinfo {pages} {950} (\bibinfo {year} {2016})}\BibitemShut
  {NoStop}%
\bibitem [{\citenamefont {Gamayun}\ \emph {et~al.}(2018)\citenamefont
  {Gamayun}, \citenamefont {Ostroukh}, \citenamefont {Gnezdilov}, \citenamefont
  {Adagideli},\ and\ \citenamefont {Beenakker}}]{Gamayun}%
  \BibitemOpen
  \bibfield  {author} {\bibinfo {author} {\bibfnamefont {O.~V.}\ \bibnamefont
  {Gamayun}}, \bibinfo {author} {\bibfnamefont {V.~P.}\ \bibnamefont
  {Ostroukh}}, \bibinfo {author} {\bibfnamefont {N.~V.}\ \bibnamefont
  {Gnezdilov}}, \bibinfo {author} {\bibfnamefont {{\.{I}}.}~\bibnamefont
  {Adagideli}},\ and\ \bibinfo {author} {\bibfnamefont {C.~W.~J.}\ \bibnamefont
  {Beenakker}},\ }\bibfield  {title} {\bibinfo {title} {Valley-momentum locking
  in a graphene superlattice with y-shaped kekul{\'{e}} bond texture},\ }\href
  {https://doi.org/10.1088/1367-2630/aaa7e5} {\bibfield  {journal} {\bibinfo
  {journal} {New Journal of Physics}\ }\textbf {\bibinfo {volume} {20}},\
  \bibinfo {pages} {023016} (\bibinfo {year} {2018})}\BibitemShut {NoStop}%
\bibitem [{\citenamefont {Andrade}\ \emph {et~al.}(2019)\citenamefont
  {Andrade}, \citenamefont {Carrillo-Bastos},\ and\ \citenamefont
  {Naumis}}]{Elias_2019}%
  \BibitemOpen
  \bibfield  {author} {\bibinfo {author} {\bibfnamefont {E.}~\bibnamefont
  {Andrade}}, \bibinfo {author} {\bibfnamefont {R.}~\bibnamefont
  {Carrillo-Bastos}},\ and\ \bibinfo {author} {\bibfnamefont {G.~G.}\
  \bibnamefont {Naumis}},\ }\bibfield  {title} {\bibinfo {title} {Valley
  engineering by strain in kekul\'e-distorted graphene},\ }\href
  {https://doi.org/10.1103/PhysRevB.99.035411} {\bibfield  {journal} {\bibinfo
  {journal} {Phys. Rev. B}\ }\textbf {\bibinfo {volume} {99}},\ \bibinfo
  {pages} {035411} (\bibinfo {year} {2019})}\BibitemShut {NoStop}%
\bibitem [{\citenamefont {Ruiz-Tijerina}\ \emph {et~al.}(2019)\citenamefont
  {Ruiz-Tijerina}, \citenamefont {Andrade}, \citenamefont {Carrillo-Bastos},
  \citenamefont {Mireles},\ and\ \citenamefont {Naumis}}]{Tijerina_2019}%
  \BibitemOpen
  \bibfield  {author} {\bibinfo {author} {\bibfnamefont {D.~A.}\ \bibnamefont
  {Ruiz-Tijerina}}, \bibinfo {author} {\bibfnamefont {E.}~\bibnamefont
  {Andrade}}, \bibinfo {author} {\bibfnamefont {R.}~\bibnamefont
  {Carrillo-Bastos}}, \bibinfo {author} {\bibfnamefont {F.}~\bibnamefont
  {Mireles}},\ and\ \bibinfo {author} {\bibfnamefont {G.~G.}\ \bibnamefont
  {Naumis}},\ }\bibfield  {title} {\bibinfo {title} {Multiflavor dirac fermions
  in kekul\'e-distorted graphene bilayers},\ }\href
  {https://doi.org/10.1103/PhysRevB.100.075431} {\bibfield  {journal} {\bibinfo
   {journal} {Phys. Rev. B}\ }\textbf {\bibinfo {volume} {100}},\ \bibinfo
  {pages} {075431} (\bibinfo {year} {2019})}\BibitemShut {NoStop}%
\bibitem [{\citenamefont {Roy}\ and\ \citenamefont
  {Herbut}(2010)}]{Bitan_2010}%
  \BibitemOpen
  \bibfield  {author} {\bibinfo {author} {\bibfnamefont {B.}~\bibnamefont
  {Roy}}\ and\ \bibinfo {author} {\bibfnamefont {I.~F.}\ \bibnamefont
  {Herbut}},\ }\bibfield  {title} {\bibinfo {title} {Unconventional
  superconductivity on honeycomb lattice: Theory of kekule order parameter},\
  }\href {https://doi.org/10.1103/PhysRevB.82.035429} {\bibfield  {journal}
  {\bibinfo  {journal} {Phys. Rev. B}\ }\textbf {\bibinfo {volume} {82}},\
  \bibinfo {pages} {035429} (\bibinfo {year} {2010})}\BibitemShut {NoStop}%
\bibitem [{\citenamefont {Po}\ \emph {et~al.}(2018)\citenamefont {Po},
  \citenamefont {Zou}, \citenamefont {Vishwanath},\ and\ \citenamefont
  {Senthil}}]{Hoi_2019}%
  \BibitemOpen
  \bibfield  {author} {\bibinfo {author} {\bibfnamefont {H.~C.}\ \bibnamefont
  {Po}}, \bibinfo {author} {\bibfnamefont {L.}~\bibnamefont {Zou}}, \bibinfo
  {author} {\bibfnamefont {A.}~\bibnamefont {Vishwanath}},\ and\ \bibinfo
  {author} {\bibfnamefont {T.}~\bibnamefont {Senthil}},\ }\bibfield  {title}
  {\bibinfo {title} {Origin of mott insulating behavior and superconductivity
  in twisted bilayer graphene},\ }\href
  {https://doi.org/10.1103/PhysRevX.8.031089} {\bibfield  {journal} {\bibinfo
  {journal} {Phys. Rev. X}\ }\textbf {\bibinfo {volume} {8}},\ \bibinfo {pages}
  {031089} (\bibinfo {year} {2018})}\BibitemShut {NoStop}%
\bibitem [{\citenamefont {Gao}\ \emph {et~al.}(2019)\citenamefont {Gao},
  \citenamefont {Torrent}, \citenamefont {Cervera}, \citenamefont {San-Jose},
  \citenamefont {S\'anchez-Dehesa},\ and\ \citenamefont
  {Christensen}}]{Penglin_2019}%
  \BibitemOpen
  \bibfield  {author} {\bibinfo {author} {\bibfnamefont {P.}~\bibnamefont
  {Gao}}, \bibinfo {author} {\bibfnamefont {D.}~\bibnamefont {Torrent}},
  \bibinfo {author} {\bibfnamefont {F.}~\bibnamefont {Cervera}}, \bibinfo
  {author} {\bibfnamefont {P.}~\bibnamefont {San-Jose}}, \bibinfo {author}
  {\bibfnamefont {J.}~\bibnamefont {S\'anchez-Dehesa}},\ and\ \bibinfo {author}
  {\bibfnamefont {J.}~\bibnamefont {Christensen}},\ }\bibfield  {title}
  {\bibinfo {title} {Majorana-like zero modes in kekul\'e distorted sonic
  lattices},\ }\href {https://doi.org/10.1103/PhysRevLett.123.196601}
  {\bibfield  {journal} {\bibinfo  {journal} {Phys. Rev. Lett.}\ }\textbf
  {\bibinfo {volume} {123}},\ \bibinfo {pages} {196601} (\bibinfo {year}
  {2019})}\BibitemShut {NoStop}%
\bibitem [{\citenamefont {Cerda-M\'endez}\ \emph {et~al.}(2013)\citenamefont
  {Cerda-M\'endez}, \citenamefont {Sarkar}, \citenamefont {Krizhanovskii},
  \citenamefont {Gavrilov}, \citenamefont {Biermann}, \citenamefont
  {Skolnick},\ and\ \citenamefont {Santos}}]{Cerda2013}%
  \BibitemOpen
  \bibfield  {author} {\bibinfo {author} {\bibfnamefont {E.~A.}\ \bibnamefont
  {Cerda-M\'endez}}, \bibinfo {author} {\bibfnamefont {D.}~\bibnamefont
  {Sarkar}}, \bibinfo {author} {\bibfnamefont {D.~N.}\ \bibnamefont
  {Krizhanovskii}}, \bibinfo {author} {\bibfnamefont {S.~S.}\ \bibnamefont
  {Gavrilov}}, \bibinfo {author} {\bibfnamefont {K.}~\bibnamefont {Biermann}},
  \bibinfo {author} {\bibfnamefont {M.~S.}\ \bibnamefont {Skolnick}},\ and\
  \bibinfo {author} {\bibfnamefont {P.~V.}\ \bibnamefont {Santos}},\ }\bibfield
   {title} {\bibinfo {title} {Exciton-polariton gap solitons in two-dimensional
  lattices},\ }\href {https://doi.org/10.1103/PhysRevLett.111.146401}
  {\bibfield  {journal} {\bibinfo  {journal} {Phys. Rev. Lett.}\ }\textbf
  {\bibinfo {volume} {111}},\ \bibinfo {pages} {146401} (\bibinfo {year}
  {2013})}\BibitemShut {NoStop}%
\bibitem [{\citenamefont {Chen}\ \emph {et~al.}(2018)\citenamefont {Chen},
  \citenamefont {Ding}, \citenamefont {Qin}, \citenamefont {He}, \citenamefont
  {Luo}, \citenamefont {Chen}, \citenamefont {Liu}, \citenamefont {Wang},
  \citenamefont {Zhang}, \citenamefont {Li}, \citenamefont {You}, \citenamefont
  {Wang}, \citenamefont {Wang}, \citenamefont {Sanders}, \citenamefont {Lu},\
  and\ \citenamefont {Pan}}]{Photonic_2019}%
  \BibitemOpen
  \bibfield  {author} {\bibinfo {author} {\bibfnamefont {C.}~\bibnamefont
  {Chen}}, \bibinfo {author} {\bibfnamefont {X.}~\bibnamefont {Ding}}, \bibinfo
  {author} {\bibfnamefont {J.}~\bibnamefont {Qin}}, \bibinfo {author}
  {\bibfnamefont {Y.}~\bibnamefont {He}}, \bibinfo {author} {\bibfnamefont
  {Y.-H.}\ \bibnamefont {Luo}}, \bibinfo {author} {\bibfnamefont {M.-C.}\
  \bibnamefont {Chen}}, \bibinfo {author} {\bibfnamefont {C.}~\bibnamefont
  {Liu}}, \bibinfo {author} {\bibfnamefont {X.-L.}\ \bibnamefont {Wang}},
  \bibinfo {author} {\bibfnamefont {W.-J.}\ \bibnamefont {Zhang}}, \bibinfo
  {author} {\bibfnamefont {H.}~\bibnamefont {Li}}, \bibinfo {author}
  {\bibfnamefont {L.-X.}\ \bibnamefont {You}}, \bibinfo {author} {\bibfnamefont
  {Z.}~\bibnamefont {Wang}}, \bibinfo {author} {\bibfnamefont {D.-W.}\
  \bibnamefont {Wang}}, \bibinfo {author} {\bibfnamefont {B.~C.}\ \bibnamefont
  {Sanders}}, \bibinfo {author} {\bibfnamefont {C.-Y.}\ \bibnamefont {Lu}},\
  and\ \bibinfo {author} {\bibfnamefont {J.-W.}\ \bibnamefont {Pan}},\
  }\bibfield  {title} {\bibinfo {title} {Observation of topologically protected
  edge states in a photonic two-dimensional quantum walk},\ }\href
  {https://doi.org/10.1103/PhysRevLett.121.100502} {\bibfield  {journal}
  {\bibinfo  {journal} {Phys. Rev. Lett.}\ }\textbf {\bibinfo {volume} {121}},\
  \bibinfo {pages} {100502} (\bibinfo {year} {2018})}\BibitemShut {NoStop}%
\bibitem [{\citenamefont {Rajagopal}\ \emph {et~al.}(2019)\citenamefont
  {Rajagopal}, \citenamefont {Shimasaki}, \citenamefont {Dotti}, \citenamefont
  {Ra\ifmmode \check{c}\else \v{c}\fi{}i\ifmmode~\bar{u}\else \={u}\fi{}nas},
  \citenamefont {Senaratne}, \citenamefont {Anisimovas}, \citenamefont
  {Eckardt},\ and\ \citenamefont {Weld}}]{Atomic_2019}%
  \BibitemOpen
  \bibfield  {author} {\bibinfo {author} {\bibfnamefont {S.~V.}\ \bibnamefont
  {Rajagopal}}, \bibinfo {author} {\bibfnamefont {T.}~\bibnamefont
  {Shimasaki}}, \bibinfo {author} {\bibfnamefont {P.}~\bibnamefont {Dotti}},
  \bibinfo {author} {\bibfnamefont {M.}~\bibnamefont {Ra\ifmmode \check{c}\else
  \v{c}\fi{}i\ifmmode~\bar{u}\else \={u}\fi{}nas}}, \bibinfo {author}
  {\bibfnamefont {R.}~\bibnamefont {Senaratne}}, \bibinfo {author}
  {\bibfnamefont {E.}~\bibnamefont {Anisimovas}}, \bibinfo {author}
  {\bibfnamefont {A.}~\bibnamefont {Eckardt}},\ and\ \bibinfo {author}
  {\bibfnamefont {D.~M.}\ \bibnamefont {Weld}},\ }\bibfield  {title} {\bibinfo
  {title} {Phasonic spectroscopy of a quantum gas in a quasicrystalline
  lattice},\ }\href {https://doi.org/10.1103/PhysRevLett.123.223201} {\bibfield
   {journal} {\bibinfo  {journal} {Phys. Rev. Lett.}\ }\textbf {\bibinfo
  {volume} {123}},\ \bibinfo {pages} {223201} (\bibinfo {year}
  {2019})}\BibitemShut {NoStop}%
\bibitem [{\citenamefont {{Naumis}}\ \emph {et~al.}(2017)\citenamefont
  {{Naumis}}, \citenamefont {{Barraza-Lopez}}, \citenamefont {{Oliva-Leyva}},\
  and\ \citenamefont {{Terrones}}}]{NaumisReview}%
  \BibitemOpen
  \bibfield  {author} {\bibinfo {author} {\bibfnamefont {G.~G.}\ \bibnamefont
  {{Naumis}}}, \bibinfo {author} {\bibfnamefont {S.}~\bibnamefont
  {{Barraza-Lopez}}}, \bibinfo {author} {\bibfnamefont {M.}~\bibnamefont
  {{Oliva-Leyva}}},\ and\ \bibinfo {author} {\bibfnamefont {H.}~\bibnamefont
  {{Terrones}}},\ }\bibfield  {title} {\bibinfo {title} {{Electronic and
  optical properties of strained graphene and other strained 2D materials: a
  review}},\ }\href {https://doi.org/10.1088/1361-6633/aa74ef} {\bibfield
  {journal} {\bibinfo  {journal} {Reports on Progress in Physics}\ }\textbf
  {\bibinfo {volume} {80}},\ \bibinfo {eid} {096501} (\bibinfo {year}
  {2017})}\BibitemShut {NoStop}%
\bibitem [{\citenamefont {Castro~Neto}\ \emph {et~al.}(2009)\citenamefont
  {Castro~Neto}, \citenamefont {Guinea}, \citenamefont {Peres}, \citenamefont
  {Novoselov},\ and\ \citenamefont {Geim}}]{Castro1}%
  \BibitemOpen
  \bibfield  {author} {\bibinfo {author} {\bibfnamefont {A.~H.}\ \bibnamefont
  {Castro~Neto}}, \bibinfo {author} {\bibfnamefont {F.}~\bibnamefont {Guinea}},
  \bibinfo {author} {\bibfnamefont {N.~M.~R.}\ \bibnamefont {Peres}}, \bibinfo
  {author} {\bibfnamefont {K.~S.}\ \bibnamefont {Novoselov}},\ and\ \bibinfo
  {author} {\bibfnamefont {A.~K.}\ \bibnamefont {Geim}},\ }\bibfield  {title}
  {\bibinfo {title} {The electronic properties of graphene},\ }\href
  {https://doi.org/10.1103/RevModPhys.81.109} {\bibfield  {journal} {\bibinfo
  {journal} {Rev. Mod. Phys.}\ }\textbf {\bibinfo {volume} {81}},\ \bibinfo
  {pages} {109} (\bibinfo {year} {2009})}\BibitemShut {NoStop}%
\bibitem [{\citenamefont {Naumis}\ \emph {et~al.}(2009)\citenamefont {Naumis},
  \citenamefont {Terrones}, \citenamefont {Terrones},\ and\ \citenamefont
  {Gaggero-Sager}}]{NaumisTerrones2009}%
  \BibitemOpen
  \bibfield  {author} {\bibinfo {author} {\bibfnamefont {G.~G.}\ \bibnamefont
  {Naumis}}, \bibinfo {author} {\bibfnamefont {M.}~\bibnamefont {Terrones}},
  \bibinfo {author} {\bibfnamefont {H.}~\bibnamefont {Terrones}},\ and\
  \bibinfo {author} {\bibfnamefont {L.~M.}\ \bibnamefont {Gaggero-Sager}},\
  }\bibfield  {title} {\bibinfo {title} {{Design of graphene electronic devices
  using nanoribbons of different widths}},\ }\href
  {https://doi.org/10.1063/1.3257731} {\bibfield  {journal} {\bibinfo
  {journal} {Appl. Phys. Lett.}\ }\textbf {\bibinfo {volume} {95}},\ \bibinfo
  {eid} {182104} (\bibinfo {year} {2009})}\BibitemShut {NoStop}%
\bibitem [{\citenamefont {Das~Sarma}\ \emph {et~al.}(2011)\citenamefont
  {Das~Sarma}, \citenamefont {Adam}, \citenamefont {Hwang},\ and\ \citenamefont
  {Rossi}}]{Sarma}%
  \BibitemOpen
  \bibfield  {author} {\bibinfo {author} {\bibfnamefont {S.}~\bibnamefont
  {Das~Sarma}}, \bibinfo {author} {\bibfnamefont {S.}~\bibnamefont {Adam}},
  \bibinfo {author} {\bibfnamefont {E.~H.}\ \bibnamefont {Hwang}},\ and\
  \bibinfo {author} {\bibfnamefont {E.}~\bibnamefont {Rossi}},\ }\bibfield
  {title} {\bibinfo {title} {Electronic transport in two-dimensional
  graphene},\ }\href {https://doi.org/10.1103/RevModPhys.83.407} {\bibfield
  {journal} {\bibinfo  {journal} {Rev. Mod. Phys.}\ }\textbf {\bibinfo {volume}
  {83}},\ \bibinfo {pages} {407} (\bibinfo {year} {2011})}\BibitemShut
  {NoStop}%
\bibitem [{\citenamefont {Yan}\ \emph {et~al.}(2015)\citenamefont {Yan},
  \citenamefont {Cruz}, \citenamefont {Barraza-Lopez},\ and\ \citenamefont
  {Yang}}]{Yan2015b}%
  \BibitemOpen
  \bibfield  {author} {\bibinfo {author} {\bibfnamefont {J.-A.}\ \bibnamefont
  {Yan}}, \bibinfo {author} {\bibfnamefont {M.~A.~D.}\ \bibnamefont {Cruz}},
  \bibinfo {author} {\bibfnamefont {S.}~\bibnamefont {Barraza-Lopez}},\ and\
  \bibinfo {author} {\bibfnamefont {L.}~\bibnamefont {Yang}},\ }\bibfield
  {title} {\bibinfo {title} {Strain-tunable topological quantum phase
  transition in buckled honeycomb lattices},\ }\href
  {https://doi.org/http://dx.doi.org/10.1063/1.4919885} {\bibfield  {journal}
  {\bibinfo  {journal} {Appl. Phys. Lett.}\ }\textbf {\bibinfo {volume}
  {106}},\ \bibinfo {pages} {183107} (\bibinfo {year} {2015})}\BibitemShut
  {NoStop}%
\bibitem [{\citenamefont {Roman-Taboada}\ and\ \citenamefont
  {Naumis}(2014)}]{Roman2015}%
  \BibitemOpen
  \bibfield  {author} {\bibinfo {author} {\bibfnamefont {P.}~\bibnamefont
  {Roman-Taboada}}\ and\ \bibinfo {author} {\bibfnamefont {G.~G.}\ \bibnamefont
  {Naumis}},\ }\bibfield  {title} {\bibinfo {title} {Spectral butterfly, mixed
  dirac-schr\"odinger fermion behavior, and topological states in armchair
  uniaxial strained graphene},\ }\href
  {https://doi.org/10.1103/PhysRevB.90.195435} {\bibfield  {journal} {\bibinfo
  {journal} {Phys. Rev. B}\ }\textbf {\bibinfo {volume} {90}},\ \bibinfo
  {pages} {195435} (\bibinfo {year} {2014})}\BibitemShut {NoStop}%
\bibitem [{\citenamefont {Roman-Taboada}\ and\ \citenamefont
  {Naumis}(2017{\natexlab{a}})}]{Taboada2017}%
  \BibitemOpen
  \bibfield  {author} {\bibinfo {author} {\bibfnamefont {P.}~\bibnamefont
  {Roman-Taboada}}\ and\ \bibinfo {author} {\bibfnamefont {G.~G.}\ \bibnamefont
  {Naumis}},\ }\bibfield  {title} {\bibinfo {title} {Topological edge states on
  time-periodically strained armchair graphene nanoribbons},\ }\href
  {https://doi.org/10.1103/PhysRevB.96.155435} {\bibfield  {journal} {\bibinfo
  {journal} {Phys. Rev. B}\ }\textbf {\bibinfo {volume} {96}},\ \bibinfo
  {pages} {155435} (\bibinfo {year} {2017}{\natexlab{a}})}\BibitemShut
  {NoStop}%
\bibitem [{\citenamefont {Roman-Taboada}\ and\ \citenamefont
  {Naumis}(2017{\natexlab{b}})}]{RomanJCP2017}%
  \BibitemOpen
  \bibfield  {author} {\bibinfo {author} {\bibfnamefont {P.}~\bibnamefont
  {Roman-Taboada}}\ and\ \bibinfo {author} {\bibfnamefont {G.~G.}\ \bibnamefont
  {Naumis}},\ }\bibfield  {title} {\bibinfo {title} {Topological phase-diagram
  of time-periodically rippled zigzag graphene nanoribbons},\ }\href
  {https://doi.org/10.1088/2399-6528/aa98fd} {\bibfield  {journal} {\bibinfo
  {journal} {Journal of Physics Communications}\ }\textbf {\bibinfo {volume}
  {1}},\ \bibinfo {pages} {055023} (\bibinfo {year}
  {2017}{\natexlab{b}})}\BibitemShut {NoStop}%
\bibitem [{\citenamefont {Fulde}(1995)}]{Fulde}%
  \BibitemOpen
  \bibfield  {author} {\bibinfo {author} {\bibfnamefont {P.}~\bibnamefont
  {Fulde}},\ }\href@noop {} {\emph {\bibinfo {title} {Electron Correlations in
  Molecules and Solids}}},\ \bibinfo {edition} {3rd}\ ed.\ (\bibinfo
  {publisher} {Springer},\ \bibinfo {year} {1995})\BibitemShut {NoStop}%
\bibitem [{\citenamefont {Wang}\ \emph {et~al.}(2017)\citenamefont {Wang},
  \citenamefont {Chen}, \citenamefont {Bomantara}, \citenamefont {Gong},\ and\
  \citenamefont {Xing}}]{TimeSSHModel}%
  \BibitemOpen
  \bibfield  {author} {\bibinfo {author} {\bibfnamefont {H.-Q.}\ \bibnamefont
  {Wang}}, \bibinfo {author} {\bibfnamefont {M.~N.}\ \bibnamefont {Chen}},
  \bibinfo {author} {\bibfnamefont {R.~W.}\ \bibnamefont {Bomantara}}, \bibinfo
  {author} {\bibfnamefont {J.}~\bibnamefont {Gong}},\ and\ \bibinfo {author}
  {\bibfnamefont {D.~Y.}\ \bibnamefont {Xing}},\ }\bibfield  {title} {\bibinfo
  {title} {Line nodes and surface majorana flat bands in static and kicked
  $p$-wave superconducting harper model},\ }\href
  {https://doi.org/10.1103/PhysRevB.95.075136} {\bibfield  {journal} {\bibinfo
  {journal} {Phys. Rev. B}\ }\textbf {\bibinfo {volume} {95}},\ \bibinfo
  {pages} {075136} (\bibinfo {year} {2017})}\BibitemShut {NoStop}%
\bibitem [{\citenamefont {Ziegler}(2007)}]{Ziegler}%
  \BibitemOpen
  \bibfield  {author} {\bibinfo {author} {\bibfnamefont {K.}~\bibnamefont
  {Ziegler}},\ }\bibfield  {title} {\bibinfo {title} {Minimal conductivity of
  graphene: Nonuniversal values from the kubo formula},\ }\href
  {https://doi.org/10.1103/PhysRevB.75.233407} {\bibfield  {journal} {\bibinfo
  {journal} {Phys. Rev. B}\ }\textbf {\bibinfo {volume} {75}},\ \bibinfo
  {pages} {233407} (\bibinfo {year} {2007})}\BibitemShut {NoStop}%
\bibitem [{\citenamefont {Foa-Torres}\ \emph {et~al.}(2014)\citenamefont
  {Foa-Torres}, \citenamefont {Roche},\ and\ \citenamefont {Charlier}}]{Foa}%
  \BibitemOpen
  \bibfield  {author} {\bibinfo {author} {\bibfnamefont {L.~E.}\ \bibnamefont
  {Foa-Torres}}, \bibinfo {author} {\bibfnamefont {S.}~\bibnamefont {Roche}},\
  and\ \bibinfo {author} {\bibfnamefont {J.-C.}\ \bibnamefont {Charlier}},\
  }\href@noop {} {\emph {\bibinfo {title} {Introduction to Graphene-Based
  Nanomaterials}}},\ \bibinfo {edition} {1st}\ ed.\ (\bibinfo  {publisher}
  {Cambridge University Press},\ \bibinfo {year} {2014})\BibitemShut {NoStop}%
\bibitem [{\citenamefont {Mak}\ \emph {et~al.}(2008)\citenamefont {Mak},
  \citenamefont {Sfeir}, \citenamefont {Wu}, \citenamefont {Lui}, \citenamefont
  {Misewich},\ and\ \citenamefont {Heinz}}]{Mak}%
  \BibitemOpen
  \bibfield  {author} {\bibinfo {author} {\bibfnamefont {K.~F.}\ \bibnamefont
  {Mak}}, \bibinfo {author} {\bibfnamefont {M.~Y.}\ \bibnamefont {Sfeir}},
  \bibinfo {author} {\bibfnamefont {Y.}~\bibnamefont {Wu}}, \bibinfo {author}
  {\bibfnamefont {C.~H.}\ \bibnamefont {Lui}}, \bibinfo {author} {\bibfnamefont
  {J.~A.}\ \bibnamefont {Misewich}},\ and\ \bibinfo {author} {\bibfnamefont
  {T.~F.}\ \bibnamefont {Heinz}},\ }\bibfield  {title} {\bibinfo {title}
  {Measurement of the optical conductivity of graphene},\ }\href
  {https://doi.org/10.1103/PhysRevLett.101.196405} {\bibfield  {journal}
  {\bibinfo  {journal} {Phys. Rev. Lett.}\ }\textbf {\bibinfo {volume} {101}},\
  \bibinfo {pages} {196405} (\bibinfo {year} {2008})}\BibitemShut {NoStop}%
\bibitem [{\citenamefont {Ryu}\ \emph {et~al.}(2007)\citenamefont {Ryu},
  \citenamefont {Mudry}, \citenamefont {Furusaki},\ and\ \citenamefont
  {Ludwig}}]{Ryu}%
  \BibitemOpen
  \bibfield  {author} {\bibinfo {author} {\bibfnamefont {S.}~\bibnamefont
  {Ryu}}, \bibinfo {author} {\bibfnamefont {C.}~\bibnamefont {Mudry}}, \bibinfo
  {author} {\bibfnamefont {A.}~\bibnamefont {Furusaki}},\ and\ \bibinfo
  {author} {\bibfnamefont {A.~W.~W.}\ \bibnamefont {Ludwig}},\ }\bibfield
  {title} {\bibinfo {title} {Landauer conductance and twisted boundary
  conditions for dirac fermions in two space dimensions},\ }\href
  {https://doi.org/10.1103/PhysRevB.75.205344} {\bibfield  {journal} {\bibinfo
  {journal} {Phys. Rev. B}\ }\textbf {\bibinfo {volume} {75}},\ \bibinfo
  {pages} {205344} (\bibinfo {year} {2007})}\BibitemShut {NoStop}%
\bibitem [{\citenamefont {Koshino}\ and\ \citenamefont
  {Ando}(2008)}]{Koshino2008}%
  \BibitemOpen
  \bibfield  {author} {\bibinfo {author} {\bibfnamefont {M.}~\bibnamefont
  {Koshino}}\ and\ \bibinfo {author} {\bibfnamefont {T.}~\bibnamefont {Ando}},\
  }\bibfield  {title} {\bibinfo {title} {Magneto-optical properties of
  multilayer graphene},\ }\href {https://doi.org/10.1103/PhysRevB.77.115313}
  {\bibfield  {journal} {\bibinfo  {journal} {Phys. Rev. B}\ }\textbf {\bibinfo
  {volume} {77}},\ \bibinfo {pages} {115313} (\bibinfo {year}
  {2008})}\BibitemShut {NoStop}%
\bibitem [{\citenamefont {Gusynin}\ \emph {et~al.}(2006)\citenamefont
  {Gusynin}, \citenamefont {Sharapov},\ and\ \citenamefont
  {Carbotte}}]{Gusynin}%
  \BibitemOpen
  \bibfield  {author} {\bibinfo {author} {\bibfnamefont {V.~P.}\ \bibnamefont
  {Gusynin}}, \bibinfo {author} {\bibfnamefont {S.~G.}\ \bibnamefont
  {Sharapov}},\ and\ \bibinfo {author} {\bibfnamefont {J.~P.}\ \bibnamefont
  {Carbotte}},\ }\bibfield  {title} {\bibinfo {title} {Unusual microwave
  response of dirac quasiparticles in graphene},\ }\href
  {https://doi.org/10.1103/PhysRevLett.96.256802} {\bibfield  {journal}
  {\bibinfo  {journal} {Phys. Rev. Lett.}\ }\textbf {\bibinfo {volume} {96}},\
  \bibinfo {pages} {256802} (\bibinfo {year} {2006})}\BibitemShut {NoStop}%
\bibitem [{\citenamefont {Herrera}\ and\ \citenamefont
  {Naumis}(2019)}]{Borophene}%
  \BibitemOpen
  \bibfield  {author} {\bibinfo {author} {\bibfnamefont {S.~A.}\ \bibnamefont
  {Herrera}}\ and\ \bibinfo {author} {\bibfnamefont {G.~G.}\ \bibnamefont
  {Naumis}},\ }\bibfield  {title} {\bibinfo {title} {Kubo conductivity for
  anisotropic tilted dirac semimetals and its application to 8-$pmmn$
  borophene: Role of frequency, temperature, and scattering limits},\ }\href
  {https://doi.org/10.1103/PhysRevB.100.195420} {\bibfield  {journal} {\bibinfo
   {journal} {Phys. Rev. B}\ }\textbf {\bibinfo {volume} {100}},\ \bibinfo
  {pages} {195420} (\bibinfo {year} {2019})}\BibitemShut {NoStop}%
\bibitem [{\citenamefont {Falkovsky}\ and\ \citenamefont
  {Pershoguba}(2007)}]{Falkovsky}%
  \BibitemOpen
  \bibfield  {author} {\bibinfo {author} {\bibfnamefont {L.~A.}\ \bibnamefont
  {Falkovsky}}\ and\ \bibinfo {author} {\bibfnamefont {S.~S.}\ \bibnamefont
  {Pershoguba}},\ }\bibfield  {title} {\bibinfo {title} {Optical far-infrared
  properties of a graphene monolayer and multilayer},\ }\href
  {https://doi.org/10.1103/PhysRevB.76.153410} {\bibfield  {journal} {\bibinfo
  {journal} {Phys. Rev. B}\ }\textbf {\bibinfo {volume} {76}},\ \bibinfo
  {pages} {153410} (\bibinfo {year} {2007})}\BibitemShut {NoStop}%
\bibitem [{\citenamefont {Kuzmenko}\ \emph {et~al.}(2008)\citenamefont
  {Kuzmenko}, \citenamefont {van Heumen}, \citenamefont {Carbone},\ and\
  \citenamefont {van~der Marel}}]{Kuzmenko}%
  \BibitemOpen
  \bibfield  {author} {\bibinfo {author} {\bibfnamefont {A.~B.}\ \bibnamefont
  {Kuzmenko}}, \bibinfo {author} {\bibfnamefont {E.}~\bibnamefont {van
  Heumen}}, \bibinfo {author} {\bibfnamefont {F.}~\bibnamefont {Carbone}},\
  and\ \bibinfo {author} {\bibfnamefont {D.}~\bibnamefont {van~der Marel}},\
  }\bibfield  {title} {\bibinfo {title} {Universal optical conductance of
  graphite},\ }\href {https://doi.org/10.1103/PhysRevLett.100.117401}
  {\bibfield  {journal} {\bibinfo  {journal} {Phys. Rev. Lett.}\ }\textbf
  {\bibinfo {volume} {100}},\ \bibinfo {pages} {117401} (\bibinfo {year}
  {2008})}\BibitemShut {NoStop}%
\bibitem [{\citenamefont {Horng}\ \emph {et~al.}(2011)\citenamefont {Horng},
  \citenamefont {Chen}, \citenamefont {Geng}, \citenamefont {Girit},
  \citenamefont {Zhang}, \citenamefont {Hao}, \citenamefont {Bechtel},
  \citenamefont {Martin}, \citenamefont {Zettl}, \citenamefont {Crommie},
  \citenamefont {Shen},\ and\ \citenamefont {Wang}}]{Horng}%
  \BibitemOpen
  \bibfield  {author} {\bibinfo {author} {\bibfnamefont {J.}~\bibnamefont
  {Horng}}, \bibinfo {author} {\bibfnamefont {C.-F.}\ \bibnamefont {Chen}},
  \bibinfo {author} {\bibfnamefont {B.}~\bibnamefont {Geng}}, \bibinfo {author}
  {\bibfnamefont {C.}~\bibnamefont {Girit}}, \bibinfo {author} {\bibfnamefont
  {Y.}~\bibnamefont {Zhang}}, \bibinfo {author} {\bibfnamefont
  {Z.}~\bibnamefont {Hao}}, \bibinfo {author} {\bibfnamefont {H.~A.}\
  \bibnamefont {Bechtel}}, \bibinfo {author} {\bibfnamefont {M.}~\bibnamefont
  {Martin}}, \bibinfo {author} {\bibfnamefont {A.}~\bibnamefont {Zettl}},
  \bibinfo {author} {\bibfnamefont {M.~F.}\ \bibnamefont {Crommie}}, \bibinfo
  {author} {\bibfnamefont {Y.~R.}\ \bibnamefont {Shen}},\ and\ \bibinfo
  {author} {\bibfnamefont {F.}~\bibnamefont {Wang}},\ }\bibfield  {title}
  {\bibinfo {title} {Drude conductivity of dirac fermions in graphene},\ }\href
  {https://doi.org/10.1103/PhysRevB.83.165113} {\bibfield  {journal} {\bibinfo
  {journal} {Phys. Rev. B}\ }\textbf {\bibinfo {volume} {83}},\ \bibinfo
  {pages} {165113} (\bibinfo {year} {2011})}\BibitemShut {NoStop}%
\bibitem [{\citenamefont {Li}\ \emph {et~al.}(2008)\citenamefont {Li},
  \citenamefont {Henriksen}, \citenamefont {Jiang}, \citenamefont {Hao},
  \citenamefont {Martin}, \citenamefont {Kim}, \citenamefont {Stormer},\ and\
  \citenamefont {Basov}}]{Li2008}%
  \BibitemOpen
  \bibfield  {author} {\bibinfo {author} {\bibfnamefont {Z.~Q.}\ \bibnamefont
  {Li}}, \bibinfo {author} {\bibfnamefont {E.~A.}\ \bibnamefont {Henriksen}},
  \bibinfo {author} {\bibfnamefont {Z.}~\bibnamefont {Jiang}}, \bibinfo
  {author} {\bibfnamefont {Z.}~\bibnamefont {Hao}}, \bibinfo {author}
  {\bibfnamefont {M.~C.}\ \bibnamefont {Martin}}, \bibinfo {author}
  {\bibfnamefont {P.}~\bibnamefont {Kim}}, \bibinfo {author} {\bibfnamefont
  {H.~L.}\ \bibnamefont {Stormer}},\ and\ \bibinfo {author} {\bibfnamefont
  {D.~N.}\ \bibnamefont {Basov}},\ }\bibfield  {title} {\bibinfo {title} {Dirac
  charge dynamics in graphene by infrared spectroscopy},\ }\href
  {https://doi.org/10.1038/nphys989} {\bibfield  {journal} {\bibinfo  {journal}
  {Nature Physics}\ }\textbf {\bibinfo {volume} {4}},\ \bibinfo {pages} {532}
  (\bibinfo {year} {2008})}\BibitemShut {NoStop}%
\bibitem [{\citenamefont {Katsnelson}(2007)}]{KATSNELSON200720}%
  \BibitemOpen
  \bibfield  {author} {\bibinfo {author} {\bibfnamefont {M.~I.}\ \bibnamefont
  {Katsnelson}},\ }\bibfield  {title} {\bibinfo {title} {Graphene: carbon in
  two dimensions},\ }\href
  {https://doi.org/https://doi.org/10.1016/S1369-7021(06)71788-6} {\bibfield
  {journal} {\bibinfo  {journal} {Materials Today}\ }\textbf {\bibinfo {volume}
  {10}},\ \bibinfo {pages} {20 } (\bibinfo {year} {2007})}\BibitemShut
  {NoStop}%
\end{thebibliography}%

\end{document}